\documentclass{elsart}

\usepackage{amsmath,
amscd,amssymb}
\usepackage{latexsym}

\newcommand{\ind}{\operatorname{ind}}

\newcommand{\codim}{\operatorname{codim}}
\newcommand{\dom}{\operatorname{dom}}
\newcommand{\im}{\operatorname{Im}}
\renewcommand{\ker}{\operatorname{Ker}}
\newcommand{\sprad}{\operatorname{sprad}}
\newcommand{\diag}{\operatorname{diag}}
\newcommand{\re}{\operatorname{Re}}
\def\Yn{X_{n,*}^{\,\,\,\, \bot}}
\def\Ym{X_{m,*}^{\,\,\,\, \bot}}
\def\Y0{X_{0,*}^{\,\,\,\, \bot}}

\theoremstyle{remark}

\numberwithin{thm}{section}
\numberwithin{equation}{section}
\allowdisplaybreaks

\begin{document}

\begin{frontmatter}



\title{Fredholm Differential Operators with Unbounded Coefficients
\thanksref{label1}}
\thanks[label1]{This project has been funded in part by the National Academy
of Sciences under the Collaboration in Basic Science and Engineering Program /
Twinning Program supported by Contract No. INT-0002341 from the National
Science Foundation. Also, the first author was
supported in part by the Research Board and Research Council of
the University of Missouri, and the second author was supported in part
by the KBN grant 5 P03A 027 21.
The authors thank K. Makarov and F. Gesztesy for many valuable
discussions, and R. Schnaubelt for important comments on the preliminary version of the
paper.}


\author{Yuri Latushkin}
\ead{yuri@math.missouri.edu}
\ead[url]{www.math.missouri.edu/\~{}yuri}

\address{Department of Mathematics, 
University of Missouri, Columbia, MO 65211, USA}

\author{Yuri Tomilov}
\ead{tomilov@mat.uni.torun.pl}
\address{Faculty of Mathematics and Computer Science, Nicholas Copernicus University, ul.
Chopina 12/18, 87-100 Torun, Poland}

\begin{abstract}
We prove that a first order linear differential operator ${\bf G}$ with unbounded operator coefficients is Fredholm 
on  spaces of functions on $\mathbb{R}$
with values in a reflexive Banach space if and only if the 
corresponding strongly continuous evolution family has 
exponential dichotomies on both $\mathbb{R}_+$ and
$\mathbb{R}_-$ and a pair of the ranges of the dichotomy
projections is Fredholm, and that the Fredholm index of ${\bf G}$ 
is equal to the Fredholm index of the pair. The operator ${\bf G}$ is the
generator of the evolution semigroup associated with the evolution family. In the case when the evolution family 
is the propagator of a well-posed differential
equation $u'(t)=A(t)u(t)$ with, generally, unbounded operators
 $A(t),t\in \mathbb{R}$, the operator ${\bf G}$ is a  closure of the operator $-\frac{d}{dt}+A(t)$.
Thus, this paper provides a complete infinite dimensional generalization of well-known 
finite dimensional results by K. Palmer, and by A. Ben-Artzi and I. Gohberg.

\end{abstract}

\end{frontmatter}

\section{Introduction and Main Results}
The celebrated Dichotomy Theorem 
asserts that a $d\times d$-matrix linear differential  operator
\begin{equation}\label{DifOp} G=-\frac{d}{dt}+A(t),\end{equation}
acting on a space of $d$-dimensional
 vector-functions on $\mathbb{R}$, is Fredholm if and 
only if the differential equation
$u'(t)=A(t)u(t)$, $t\in \mathbb{R}$, 
has exponential dichotomies on both $\mathbb{R}_+=[0,\infty)$ 
and $\mathbb{R}_-=(-\infty ,0]$; moreover, the
Fredholm index of $G$ is equal to the difference of the ranks 
of the dichotomies. K. Palmer proved this result in \cite{Pa,Pal} for the case when  $G$ acts  on a space of
continuous vector-functions. A. Ben-Artzi and I. Gohberg in \cite{BAG} proved this
result in the case when $G$ acts on $L_2(\mathbb{R};\mathbb{C}^d)$
and $A\in L_\infty (\mathbb{R};\mathcal{L}(\mathbb{C}^d))$. 
Also, we remark on an earlier paper by R. Sacker~\cite{S79}, 
where the "if"-part
of this result and the index formula 
were proved in the framework of linear
skew-product flows over the hull of $A$.  For further 
developments of the latter approach see  \cite {JPS,SS94,SellYou}, and the
bibliographies therein.

The Dichotomy Theorem is important in many
questions of finite dimensional dynamics. 
This theorem is instrumental in the study of spectral stability of 
travelling waves; see, e.g. \cite{Sa} and numerous references
therein. Motivated by applications to the study of partial 
differential equations, several steps have been made to generalize 
the Dichotomy Theorem for infinite dimensional
setting and unbounded operators $A(t)$. 
We mention here important results in \cite{DGL},
\cite[Thm.1.1]{HSS}, \cite{Lin}, \cite{MP,MPVL}, \cite[Thm.1]{SS2}, \cite{RR}, 
\cite[Thm.2.6]{SS1}, \cite{Z}, see also
the bibliographies in these papers, and the work of A. Baskakov 
\cite{Ba}-\cite{B}. Also, recently the infinite dimensional 
Dichotomy Theorem  gained additional importance due to
connections with infinite dimensional Morse theory, see \cite{AM1,AM,DGL,RS} and the literature therein. 
In the above mentioned papers infinite dimensional versions
of the Dichotomy Theorem have been proved either for important particular classes of operators $A(t)$, or under some additional assumptions on the solutions of the
differential equation $u'=A(t)u$ or its adjoint, or on the corresponding
evolution family (the propagator of the differential equation). These assumptions have been used to deal with the following
 principal differences between the finite dimensional and the infinite
dimensional settings: (a) Difficulties to prove the closedness of the
subspaces
of initial data that generate solutions of the equation $u'(t)=A(t)u(t)$ and,
respectively, the adjoint equation, that are bounded at $+\infty$ and,
respectively, $-\infty$ (see, e.g., \cite[Lem.7.6,7.11(A)]{SS94},
\cite[Lem.2.3]{SS2}); (b) That the propagator of the
differential equation or/and its adjoint may have a nontrivial kernel (see,
e.g., \cite[Ass.1]{Ba}, \cite[Hyp.5]{SS2}, \cite[Hyp.(U1)]{SS1}); and (c) That both stable
and unstable dichotomy subspaces for the equation might be infinite
dimensional (cf. \cite{KS,SS2,SS94,SS1,SV} and see Examples \ref{EJ1612.1} and \ref{PerShr}
below).

The main goal of the current paper is to prove an 
infinite dimensional version of the Dichotomy Theorem without any 
special restrictions on the operators $A(t)$. The corresponding differential
operator is considered on the space $L_p=L_p(\mathbb{R};X)$,
$p\in[1,\infty)$, or on $C_0(\mathbb{R};X)$, the space of continuous $X$-valued
functions vanishing at $\pm\infty$. The Banach space $X$ is 
assumed\footnote{We
suspect that one can remove the reflexivity assumption,
mainly used in Proposition \ref{PS3.1}, 
 but prefer not to pursue
this here.} to be reflexive. Both the formulation and the proof of the
Dichotomy Theorem in this unrestricted setting are quite different from the
ones known in the literature. 

To achieve this goal, as our starting
point, we consider not the differential equation $u'=A(t)u$, 
but a strongly continuous exponentially bounded evolution family 
$\{ U(t,\tau )\}_{t\geq \tau}$, $t,\tau \in \mathbb{R}$, on 
$X$. In particular, if the differential equation is well-posed 
(see Section \ref{S5} and cf. \cite[p.58]{ChLa99} and
\cite[Def.VI.9.1]{EnNa99}), 
then $U(t,\tau )$ is its propagator (Cauchy operator). A 
more important infinite dimensional issue is related
to the definition of the operator $G$ in \eqref{DifOp}. 
A quite natural first try is to define $G$, $(Gu)(t)=-u'(t)+A(t)u(t)$,
say, on $L_p$, as an operator with the domain 
\begin{equation}\label{domG}
\dom G=W_p^1
\cap \{ u\in L_p:u(t)\in \dom A(t) \,\, \text{a.e.}, 
\, A(\cdot )u(\cdot )\in L_p\},
\end{equation}
where $W_p^1=W_p^1(\mathbb{R};X)$, $p\in[1,\infty)$, is the Sobolev space so
that $W_p^1=\dom(-d/dt)$. 
This choice of $G$, however, appears 
to be unnecessarily restrictive since this operator might not be closed,
see, e.g., \cite[Sec.2(c)]{Sch}. 
To settle this issue, we consider instead a certain closed extension, ${\bf G}$, of the operator $G$. The operator ${\bf G}$ is the generator of a so
called {\em evolution semigroup} 
$\{ T^t\}_{t\geq 0}$ on $L_p$
or $C_0(\mathbb{R};X)$, see Lemma \ref{DefESG} below. Recently, the evolution semigroups and their generators
have been successfully used to characterize stability of
evolution families, and their exponential dichotomy on $\mathbb R$, 
see \cite{ChLa99,Sch} and the bibliographies therein, 
\cite{Br,BCT,NgRaSc98,Minh1}, and also \cite[Lem.IV.3.3]{DK} 
and \cite[Chap.10]{LevitZhik} for more classical 
but related approach. However, the complete characterization
of the Fredholm property of the 
generator of the evolution semigroup given in this paper
appears to be new. 
Our principal result reads as follows.

\begin{thm}\label{main} 
Assume that $\{U(t,\tau)\}_{t\ge\tau}$, $t,\tau\in\mathbb{R}$, is a strongly
continuous exponentially bounded evolution family on a reflexive
Banach space $X$, and let $\mathbf{G}$ denote the generator of the associated
evolution semigroup defined on $L_p(\mathbb{R};X)$, $p\in[1,\infty)$,
or on $C_0(\mathbb{R};X)$.
Then  
\begin{equation}\label{GFr}
\text{the operator $\mathbf{G}$ is Fredholm}
\end{equation} 
if and only if there exist $a\leq b$ in $\mathbb{R}$ 
such that the following two conditions hold:
\begin{enumerate}\item[(i)] The evolution family 
$\{ U(t,\tau)\}_{t\geq \tau}$ has exponential dichotomies 
$\{ P^-_t\}_{t\leq a}$ and $\{ P^+_t\}_{t\geq b}$ on
$(-\infty ,a]$ and $[b,\infty)$, respectively;
\item[(ii)] The {\em node operator} $N(b,a)$, acting 
from $\ker P^-_a$ to $\ker P^+_b$ by the rule 
$N(b,a)=(I-P^+_b)U (b,a)|_{\ker P^-_a}$, is
Fredholm.\end{enumerate}
Moreover, if \eqref{GFr} holds, then 
$\dim \ker {\bf G}=\dim \ker N(b,a)$, 
$\codim \im {\bf G}=\codim \im N(b,a)$, 
and $\ind {\bf G} =\ind N(b,a)$.
\end{thm}

Recall that a pair of subspaces $(W, V)$ in $X$ is called a 
{\em Fredholm pair} provided $\alpha(W,V):=\dim (W\cap V)<\infty$, 
the subspace $W+V$
is closed, and $\beta(W,V):=\codim (W+V)<\infty$; the 
{\em Fredholm index} of the pair is defined as $\ind (W,V)=\alpha(W,V)-
\beta(W,V)$, see, e.g.,
\cite[Sec.IV.4.1]{Kato}.
 Theorem~\ref{main} 
can be equivalently reformulated as follows.
\begin{thm}\label{mainRef} Under the assumptions in 
Theorem~\ref{main}, \eqref{GFr} is fulfilled if and only 
if the following two conditions hold:
\begin{enumerate}\item[(i')] The evolution family $\{ U(t,\tau )\}_{t\geq \tau}$ has exponential dichotomies $\{ P^-_t\}_{t\leq 0}$ and $\{ P^+_t\}_{t\geq 0}$ on
$\mathbb{R}_-$ and $\mathbb{R}_+$, respectively; 
\item[(ii')] The pair of subspaces
$(\ker P_0^-,\im P_0^+)$
 is Fredholm in $X$.\end{enumerate}
Moreover, if \eqref{GFr} holds, then
$\dim \ker {\bf G}=\alpha(\ker P^-_0, \im P^+_0)$, 
$\codim \im {\bf G}=\beta(\ker P^-_0,\im P^+_0)$, and $\ind {\bf G}=\ind
(\ker P^-_0, \im P^+_0)$.\end{thm}

Note that $N(0,0)=(I-P_0^+)|_{\ker P_0^-}:\ker P_0^-\to\ker P_0^+$,
and one can show that condition (ii') in Theorem \ref{mainRef} 
is equivalent to 
condition (ii) in Theorem \ref{main} with $a=0=b$, see Lemma \ref{REQ} below.

Let $J$ be  one of the intervals $\mathbb R_+$, $\mathbb R_{-}$, or
 $\mathbb R.$
 Recall that a family $\{ U(t,\tau )\}_{t\geq \tau}$, $t, \tau \in J$, 
 of bounded linear operators  on $X$  is called a
{\em strongly continuous exponentially bounded
evolution family} on  $J$ 
if: (1) for each $x\in X$ the map 
$(t,\tau )\mapsto U(t,\tau )x$ is continuous for all $t \ge \tau$ in $ J$; 
(2) for some $\omega \in \mathbb R$ the inequality $\sup \{ \| e^{-\omega (t-\tau )}U(t,\tau )\| : t, 
\tau \in J, t \ge \tau \} <\infty$ holds;
 and (3) $U(t,t)=I$, $U(t,\tau )=U(t,s)U(s,\tau )$
 for all $t\geq s\geq \tau$ in $J.$  We say that $\{
U(t,\tau )\}_{t\geq \tau}$ has {\em exponential dichotomy} 
$\{ P_t\}_{t\in J}$ on $J$ with {\it dichotomy constants} $M\geq 1$ and $\alpha >0$ if $P_t,t\in J$, 
are bounded projections on $X$, and for
 all $t\geq \tau$ in $J$  the following assertions
hold:

\begin{enumerate}\item[(i)]$U(t,\tau )P_\tau =P_tU(t,\tau )$ (intertwining property),

\item[(ii)] the restriction $U(t,\tau )|_{\ker P_\tau}$ 
of the operator $U(t,\tau )$ 
 is an invertible operator from $\ker P_\tau $
to $\ker P_t$;

\item[(iii)] the following {\em stable} and {\em unstable} dichotomy estimates hold:
\[\| U(t,\tau )|_{\im P_\tau}\| \leq Me^{-\alpha (t-\tau )}
\text{ and } \| (U(t,\tau )|_{\ker P_\tau})^{-1}\| \leq Me^{-\alpha
(t-\tau)}.\]\end{enumerate}
For the notion of exponential dichotomy we refer to the 
classical books \cite{Henry,LevitZhik},
and to newer work in \cite{ChLa99,ChLe,EnNa99,Sch,SellYou}, and
the extensive bibliographies therein. Note that (i)-(iii) imply that for every $x \in X$ the function
$ t \to P_t x$ is continuous on $J$ and $\sup_{t \in J} \| P_t \| < \infty,$ see e.g. 
\cite[Lem.4.2]{NgRaSc98} or \cite[Lem.IV.1.1,IV.3.2]{DK}.

Recall that the evolution semigroup $\{ T^t\}_{t\geq 0}$ is defined on 
$L_p(\mathbb{R};X)$, $p \in [1,  \infty)$, or on $C_0(\mathbb{R};X)$,
 by the formula $T^tu(\tau )=U(\tau,\tau -t)u(\tau-t),\tau \in \mathbb{R}$, see
\cite{ChLa99}. This is a strongly continuous semigroup, and we let
$\bf G$ denote its generator. Alternatively,
the generator {\bf G} can be described as follows (see \cite[Prop.4.32]{ChLa99}, 
and cf. \cite[Thm.1]{Bd1}, \cite[Lem.1.1]{NgRaSc98}, \cite[Lem.1.1]{Minh1}). 
\begin{lem}\label{DefESG} A function $u$ belongs to the domain
$\dom {\bf G}$ of the operator ${\bf G}$ on $L_p(\mathbb{R};X)$, 
$p \in [1,  \infty)$, resp., on $C_0(\mathbb{R};X)$,
 if and only if $u\in L_p(\mathbb{R};X)\cap C_0(\mathbb{R};X)$,
resp., $u\in C_0(\mathbb{R};X)$, 
and there exists an $f\in L_p(\mathbb{R};X)$, resp., 
$f\in C_0(\mathbb{R};X)$, such that
\begin{equation}
u(t)=U(t,\tau )u(\tau)-\int^t_\tau U(t,s)f(s)ds,\quad 
\text{for all
$t\ge\tau$ in $\mathbb{R}$}\label{DGG}.\end{equation}
If \eqref{DGG} holds, then ${\bf G}u=f$.
\end{lem}
We stress that \eqref{DGG} is a mild 
reformulation of the inhomogeneous differential    
equation $ u'(t)=A(t)u(t)+f(t)$, $t\in
\mathbb{R}$.
If $\{ U(t,\tau )\}_{t\geq \tau}$ is the propagator of a 
well-posed
 differential equation $u'(t)=A(t)u(t)$, $t \in \mathbb R$, 
with, generally, unbounded operators
$A(t)$, then the set $\dom G$ from \eqref{domG} is a core for ${\bf G}$,
see \cite[Thm.3.12]{ChLa99} and \cite[Prop.4.1]{Sch}. Thus, if the operator
$G$ with the domain $\dom G$ from \eqref{domG} is a closed operator on
$L_p(\mathbb{R};X)$, $p\in[1,\infty)$, resp., on $C_0(\mathbb{R};X)$,
then ${\bf G}=G$.

Under an {\em a priori} assumption that assertion (i) in 
Theorem~\ref{main} holds, the equivalence of \eqref{GFr} and (ii), 
and the index formula, have been studied in
\cite[Thm. 4]{BaskDAN02} and in \cite[Thm. 8]{B}. Therefore, 
in the current paper we will concentrate mostly on the main new
contribution which is a proof of the implication \eqref{GFr} 
$\Rightarrow$ (i') in Theorem \ref{mainRef}. Our strategy is to pass from the {\em
differential} operator {\bf G} on $L_p(\mathbb{R};X)$, $p\in[1,\infty)$, resp.,
on $C_0(\mathbb{R};X)$, to an associated {\em difference} 
operator, $D$, defined on the space $\ell_p(\mathbb{Z};X)$, $p\in[1,\infty)$,
resp., on the space $c_0(\mathbb{Z};X)$ of sequences vanishing at
$\pm\infty$, by the rule
\begin{equation}\label{2.1} D:(x_n)_{n\in \mathbb{Z}}
\mapsto (x_n-U(n,n-1)x_{n-1})_{n\in \mathbb{Z}}.\end{equation}
This strategy has a long history that goes back to 
D. Henry \cite[Thm.7.6.5]{Henry}. It was successfully used to treat the dichotomy on $\mathbb{R}$ and {\em
invertible} operators {\bf G}, see \cite[Thm.2]{BaskMatNotes96}, 
\cite[Thm.2]{Br},  \cite[Lem.3.3]{LMS}, \cite[Sec.5]{BAG1},
\cite[Thm.4.16,4.37]{ChLa99} (and also \cite[Thm.7.9]{ChLa99} and \cite[Thm.4.1]{PlissSell} or \cite[Thm.45.8]{SellYou} 
for a related case of linear skew-product flows on Banach spaces). 
The justification of this strategy for dichotomies on 
$\mathbb{R}_+$ and $\mathbb{R}_-$ and {\em Fredholm} operators {\bf G} 
is given in the following theorem
(cf. \cite[Thm.2]{Ba}, \cite[Thm.1]{Bd1} and \cite[Thm.2]{T91}).

\begin{thm}\label{FrGandD} Assume that $\{ U(t,\tau )\}_{t\geq \tau}$, 
$t,\tau \in \mathbb{R}$, is a strongly continuous exponentially bounded 
evolution family on a Banach space $X$, let
${\bf G}$ denote the generator of the associated evolution semigroup 
on $L_p(\mathbb{R};X)$, $p\in [1,\infty)$, resp.,
$C_0(\mathbb{R};X)$, and let $D$ be
the difference operator on $\ell_p(\mathbb{Z};X)$,
$p\in[1,\infty)$, resp., on $c_0(\mathbb{Z};X)$,
defined in \eqref{2.1}. Then $\im {\bf G}$ is closed if and only 
if $\im D$ is closed, and $\dim \ker {\bf G}=\dim \ker D$ and $\codim \im
{\bf G}=\codim \im D$. In particular, the operator {\bf G} 
is Fredholm if and only if $D$ is Fredholm, and $\ind {\bf
G}=\ind D$.\end{thm}

By the following simple lemma,  
an exponential dichotomy on $\mathbb{Z}_\pm$ extends to an
exponential dichotomy 
on $\mathbb{R}_\pm$ (cf. \cite[Ex.7.6.10]{Henry}). 
\begin{lem}\label{DCDich} Assume that $\{U(t,\tau )\}_{t\geq \tau}$, 
$t,\tau \in \mathbb{R}$, is a strongly continuous exponentially bounded 
evolution family on
a Banach space $X$. The discrete evolution family $\{U(n,m)\}_{n\geq m}$, $n,m\in \mathbb{Z}$, has an exponential dichotomy $\{ P^+_n\}_{n\geq 0}$ on
$\mathbb{Z}_+$, resp., $\{ P^-_n\}_{n\leq 0}$ on $\mathbb{Z}_-$, 
if and only if the  family $\{ U(t,\tau )\}_{t\geq \tau}$, 
$t,\tau \in \mathbb{R}$, has an
exponential dichotomy
$\{P^+_t\}_{t\geq 0}$ on
$\mathbb{R}_+$, resp., 
$\{ P^-_t\}_{t\leq 0}$ on $\mathbb{R}_-$.\end{lem}

Therefore, the assertion
\eqref{GFr} $\Rightarrow$ (i') required for the proof of 
Theorems \ref{main} and \ref{mainRef},
follows from our next  theorem (this main technical result of the current 
paper is proved in Section~\ref{FDOED}).

\begin{thm}\label{FREDSA} Assume that $X$ is a reflexive
Banach space, and the operator $D$ is Fredholm
on $\ell_p(\mathbb{Z};X)$, $p\in[1,\infty)$,
or on $c_0(\mathbb{Z};X)$. 
Then the discrete evolution family $\{
U(n,m)\}_{n\geq m}$, $n,m\in\mathbb{Z}$,
 has exponential dichotomies $\{ P^+_n\}_{n\geq
0}$ and $\{ P^-_n\}_{n\leq 0}$ on $\mathbb{Z}_+$ and
$\mathbb{Z}_-$, respectively.\end{thm}

Our strategy of the proof of Theorems \ref{main}, \ref{mainRef},
and \ref{FREDSA} is as follows.  We
will identify a family of subspaces $\{\Yn\}_{n\in \mathbb{Z}}$
in $X$ that is $U(n,m)-$invariant in the sense that
$U(n,m)(\Ym) \subseteq \Yn$ for $n\geq m$ in $\mathbb{Z}$, see Section~\ref{NP}. Next, we will show that the
restricted evolution family $\{U(n,m)|_{\Ym}\}_{n\geq m}$ has
a ``punctured'' exponential dichotomy $\{ P_n\}_{n\in \mathbb{Z}}$ on
$\mathbb{Z}$, that is, we will show the following: (1) There exist projections
$P_n$ defined on $\Yn$ that intertwine the operators $U(n,m)|_{\Ym}$ for $n\ge m>0$
and for $0\ge n\ge m$; (2) the stable and unstable dichotomy estimates hold 
for the operators $U(n,m)|_{\Ym}$ restricted
on the subspaces $\im P_m$ and $\ker P_m$; and (3) there is
a surjective {\em reduced} node operator acting from $\ker P_0$ to
$\ker P_1$. Further, we will identify a family of subspaces in $X^*$, 
the adjoint space, such that a corresponding family of restrictions of the adjoint operators
$U(n,m)^*$, $n\ge m$, enjoys similar properties for a family of projections
$\{P_{n,*}\}_{n \in \mathbb Z}$ defined on certain subspaces of $X^*$. The punctured
dichotomies just described are constructed in Section \ref{PD}.
To conclude the proof of Theorem \ref{FREDSA},
we define in Section \ref{FDOED} the dichotomies
$\{P_n^+\}_{n \ge 0}$ and $\{P_n^- \}_{n \le 0}$ using 
$\{P_n\}_{n \in \mathbb Z}$ and $\{ (P_{n,*})^* \}_{n \in \mathbb Z}$.
In Section \ref{PFMT} we finish the proof of Theorems \ref{main}
and \ref{mainRef}. This includes a proof 
(based on a new approach) of the fact
that (i) and (ii) in Theorem \ref{main} imply \eqref{GFr}, 
and the formulas for  the defect numbers
and index. Theorem \ref{FrGandD} and Lemma \ref{DCDich}
are proved in Section \ref{SFDD}.
Finally, in Section \ref{S5} we discuss several special cases when conditions of 
Theorems \ref{main} 
and \ref{mainRef} could be easily checked, and briefly mention several classes of
problems where these theorems could be applied.

\section{Notation and Preliminaries}\label{NP}
{\bf Notation.}
We denote: $\mathbb R_{+}:=\{t \in \mathbb R : t \ge 0\},$
$\mathbb R_{-}:=\{t \in \mathbb R : t \le 0\}$, $\mathbb Z_{+}:=\{n \in \mathbb Z : n \ge 0\},$
$\mathbb Z_{-}:=\{n \in \mathbb Z : n \le 0\}$,
$\mathbb{T}=\{\lambda \in\mathbb{C}:|\lambda|=1\}$; $X$ is a Banach space; 
$X^*$ is the adjoint space; $A^*$, $\dom A$, $\ker A$ and $\im A$ are the
 adjoint, domain, kernel and range of an operator $A;$ $\sigma(A), \rho(A)$ and $\sprad (A)$ denote the spectrum, 
the resolvent set, and the spectral radius
of $A;$ symbol
$A|_Y$ denotes the restriction of $A$ on
a subspace $Y\subset X$; the Banach space  of 
bounded linear operators from $X$ to $Y$ is denoted by $\mathcal{L}(X,Y)$;
a generic constant is denoted by $c$. 
We use boldface to denote sequences, e.g., ${\bf x}=(x_n)_{n\in
\mathbb{Z}},x_n\in X$. For $n\in\mathbb{Z}$ 
the $n$-th standard ort in $\ell_p(\mathbb{Z};X)$ or $c_0(\mathbb{Z};X)$ 
is denoted by ${\bf e}_n=(\delta_{nk})_{k\in \mathbb{Z}}$, 
where $\delta_{nk}$ is the
Kronecker delta. If $x\in X$ then we denote by $x\otimes {\bf e}_n=(x\delta_{nk})_{k\in \mathbb{Z}}$ the sequence $x\otimes {\bf e}_n=(x_k)_{k\in \mathbb{Z}}$ such
that $x_n=x$ and $x_k=0$ for $k\neq n$. 

For  subspaces $Y\subset
X$ and $Y_*\subset X^*$ we denote 
$Y^\bot =\{\xi \in X^*: \langle x,\xi \rangle=0$ for all $x\in
Y\}$ and $Y_*^\bot =\{x \in X: \langle x,\xi \rangle=0$ for all $\xi\in
Y_*\}$,  where $\langle \cdot ,\cdot \rangle$ is
the $(X,X^*)$-pairing. If $X=X_1\oplus X_2$, a direct sum
decomposition, then we identify $(X_1)^*=X^\bot_2$ and
$(X_2)^*=X^\bot_1$. If $P$ is a projection on
$X$, then $P^*$ is a projection on $X^*$ with
$\im P^*=(\ker P)^\bot=(\im P)^*$ and $\ker P^*=(\im P)^\bot=(\ker P)^*$.
If $(P,Q)$ is a pair of projections on $X$, then in the direct sum
decompositions $X=\im P\oplus \ker P$ and 
$X=\im Q\oplus \ker Q$ any 
operator $A$ bounded on $X$ can be written as the following $(2\times 2)$
operator matrix:
\begin{equation}\label{1B.zero}
A=\begin{bmatrix}P\\I-P\end{bmatrix} A[Q\quad I-Q]=\begin{bmatrix} PAQ
& PA(I-Q)\\(I-P)AQ & \quad (I-P)A(I-Q)\end{bmatrix}.
\end{equation}
If $AQ=PA$ then this matrix is diagonal with the diagonal entries being
$A|_{\im Q}$ and $A|_{\ker Q}$. If 
 $A(\im Q)\subseteq \im P$, or $AQ=PAQ$, then
we identify $A|_{\im Q}=AQ:\im Q\to\im P$, and write
\begin{equation}\label{1B.1} A=\begin{bmatrix}A|_{\im Q} & PA(I-Q)\\ 0
& (I-P)A|_{{\ker} Q}\end{bmatrix}.\end{equation} 
For brevity, we denote: $L_p=L_p(\mathbb{R};X)$,
$\ell_p=\ell_p(\mathbb{Z};X)$,
$\ell_{q,*}=\ell_q(\mathbb{Z};X^*)$,
$c_0=c_0(\mathbb{Z};X)$,
$c_{0,*}=c_0(\mathbb{Z};X^*)$, and remark that $(\ell_p)^*=\ell_{q,*}$
for $p\in[1,\infty)$, $q\in(1,\infty]$, $p^{-1}+q^{-1}=1$,
and $(c_0)^*=\ell_{1,*}$; if $X$ is  reflexive then $(c_{0,*})^*=\ell_1$.

{\bf Fibers of the kernel and cokernel of {\boldmath$D$\unboldmath}.}
In Sections \ref{NP}, \ref{PD}, and \ref{FDOED}
 we assume that the operator $D$ 
from \eqref{2.1} is Fredholm 
on $\ell_p(\mathbb{Z};X)$, $p\in[1,\infty)$, or on $c_0(\mathbb{Z};X)$.
Consider the operator $D^*$ adjoint of $D$:
\begin{align}
\label{2.4}D^*:(\xi_n)_{n\in \mathbb{Z}}&\mapsto
(\xi_n-U(n+1,n)^*\xi_{n+1})_{n\in \mathbb{Z}}.\end{align}
If the operator $D$ is acting on $\ell_p$, $p\in[1,\infty)$, resp.,
on $c_0$, then the adjoint operator $D^*$ is acting on $\ell_{q,*}$,
$q\in(1,\infty]$, resp., on $\ell_{1,*}$, and for  sequences
$(x_n)_{n\in\mathbb{Z}}$ and $(\xi_n)_{n\in\mathbb{Z}}$ from the 
spaces of $X$-- or $X^*$--valued sequences we have: 
\begin{align} \label{2.7} {\ker} \, D &=\{ (x_n)_{n\in \mathbb{Z}}:
x_n=U(n,m)x_m\text{ for all } n\geq m\text{ in }\mathbb{Z}\},\\
\label{2.8}{\ker} \, D^*&=\{(\xi_n)_{n\in \mathbb{Z}}:
\xi_m=U(n,m)^*\xi_n\text{ for all }n\geq m\text{ in }
\mathbb{Z}\}.\end{align} 
For each $n\in \mathbb{Z}$ we define the following subspaces:
\begin{align}\label{3.1} X_n:&=\{x\in X:\text{ there exists }
(x_k)_{k\in \mathbb{Z}}\in {\ker} D \text{ so that } x=x_n\};\\
\label{3.2}X_{n,*}:&=\{\xi \in X^*:\text{ there exists } (\xi
_k)_{k\in
\mathbb{Z}}\in {\ker} D^* \text{ so that } \xi =\xi_n\}.\end{align}

\begin{lem}\label{L4.1}  For all $n\in \mathbb Z$ and $m \in \mathbb{Z}, m \le n,$ the following
assertions hold:
\begin{enumerate}\item[(i)] $\dim X_n\leq \dim \ker D <\infty$ and $\dim
X_{n,*}\leq \dim {\ker} D^*<\infty$,\footnote{In fact,
$\dim X_n=\dim \ker D$ and $\dim
X_{n,*}= \dim {\ker} D^*$, see Corollary \ref{CS11.1}.} 
\item[(ii)] $U(n,m)X_m\subset X_n$; moreover, the operator \\
$U(n,m)|_{X_m}:X_m\to X_n$ is invertible;
\item[(iii)] $U(n,m)^*X_{n,*}\subset X_{m,*}$; moreover, the operator \\
$U(n,m)^*|_{X_{n,*}}:X_{n,*}\to X_{m,*}$ is invertible;
\item[(iv)] $U(n,m)\Ym\subset \Yn$ and $\codim \Yn=\dim X_{n,*}<\infty$;
\item[] $U(n,m)^*X_n^\bot\subset X_m^\bot$ and $\codim X_n^\bot=\dim
X_n<\infty$;
\item[(v)] $X_n\subset \Yn$ and $X_{n,*}\subset X_n^{\bot}$.
\end{enumerate}\end{lem}

\begin{pf} (i) follows from the definition of $X_n$ and $X_{n,*}$ since $D$
is Fredholm.

(ii) Fix $x\in X_m$, and pick a sequence $(x_k)_{k\in
\mathbb{Z}}\in \ker D$ such that $x=x_m$. Using \eqref{2.7}, we have
$x_n=U(n,m)x_m$. Since $(x_k)_{k\in \mathbb{Z}}\in
{\ker} D$, this shows that $U(n,m)x_m\in X_n$ by the
definition of $X_n$. Since $\dim X_n<\infty$, in order
  to show that the operator $U(n,m)|_{X_m}:X_m\to X_n$ is
invertible, it suffices to check that it is surjective.
So fix an $x\in X_n$, and pick a sequence $(x_k)_{k\in
\mathbb{Z}}\in {\ker} D$ such that $x=x_n$. Using \eqref{2.7},
we have $x_n=U(n,m)x_m$. By the definition of
$X_m$, we have $x_m\in X_m$. Thus
$x=U(n,m)x_m$ for some $x_m\in X_m$, and $U(n,m)|_{X_m}:X_m\to
X_n$ is an isomorphism.

(iii) Exactly as in (ii), using \eqref{2.8} instead of \eqref{2.7}.

(iv) For $y\in \Ym$ we have $\langle
y,\xi\rangle =0$ for all $\xi \in X_{m,*}$. If $\eta \in X_{n,*}$
then $U(n,m)^*\eta \in X_{m,*}$ by (iii) and $\langle U(n,m)y,\eta
\rangle =\langle y, U(n,m)^*\eta \rangle =0$. Thus, $U(n,m)y\in
\Yn$. The proof for $U(n,m)^*$ is similar.

(v) Fix $x\in X_n$ and $\xi \in X_{n,*}$, and pick sequences
$(x_k)_{k\in \mathbb{Z}}\in {\ker} D$ and $(\xi_k)_{k \in
\mathbb{Z}}\in \ker \, D^*$ such that $x=x_n$ and $\xi =\xi_n$.
Then
\begin{equation*}\begin{split} \infty >\sum_{k\in \mathbb{Z}}\langle
x_k,\xi_k \rangle &=\sum_{k\geq n}\langle x_k,\xi_k\rangle
+\sum_{k<n}\langle x_k,\xi_k\rangle\\
&=\sum_{k\geq n}\langle U(k,n)x_n,\xi_k\rangle +\sum_{k<n}\langle
x_k,U(n,k)^*\xi_n\rangle\\
&=\sum_{k\geq n}\langle x_n, U(k,n)^*\xi_k\rangle
+\sum_{k<n}\langle
U(n,k)x_k,\xi_n\rangle\\
&= \sum_{k\geq n}\langle x_n,\xi_n\rangle +\sum_{k<n}\langle
x_n,\xi_n\rangle =\sum_{k\in \mathbb{Z}}\langle x,\xi\rangle
,\end{split}\end{equation*}
where \eqref{2.7} and \eqref{2.8} have been used. Thus, $\langle
x,\xi\rangle=0$.\hfill$\Box$\end{pf}

{\bf Invertibility of a part of {\boldmath{$D$}\unboldmath}.}
Let $X'_n\subset \Yn$ denote any direct
complement of the finite dimensional subspace $X_n$ in $\Yn$. 
Let $Y_n$ denote any direct complement of the finite codimensional
subspace $\Yn$ in $X$.  We have the following direct sum decomposition:
\begin{equation}\label{6.1} X=\Yn\oplus Y_n=X_n\oplus X'_n\oplus
Y_n,\quad n\in \mathbb{Z}.\end{equation}
Define the following closed subspace of $\ell_p(\mathbb{Z};X)$,
$p\in[1,\infty)$, or of $c_0(\mathbb{Z};X)$:
\begin{equation}\label{6.2}\mathcal{F}:=\{(y_n)_{n\in \mathbb{Z}}:
y_n\in \Yn\text{ for each } n\in
\mathbb{Z}\}.\end{equation}

\begin{lem}\label{L6.1} Operator $D$ leaves $\mathcal{F}$ invariant,
and $D|_{\mathcal{F}}$ is surjective on $\mathcal{F}$.
\end{lem}

\begin{pf} If $y_n\in \Yn$ and $y_{n-1}\in X_{n-1,*}^{\,\,\, \bot}$ then
$y_n-U(n,n-1)y_{n-1}\in \Yn$ by Lemma~\ref{L4.1}(iv), and
$D\mathcal{F}\subset\mathcal{F}$.
To see that $D|_{\mathcal{F}}$ is surjective, we claim, 
first, that $\mathcal{F}\subset \im D$. Since $D$ is
Fredholm, its range is closed. Therefore, 
$\im D$ is the set of sequences ${\bf y}$
such that $\langle {\bf y}$, {\boldmath{$\xi$}\unboldmath}$\rangle=0$
for all sequences {\boldmath{$\xi$}\unboldmath}$\in\ker D^*$.
So, to prove the claim it suffices to show that   
{${\bf y}\bot$\boldmath{$\xi$}\unboldmath}  for all sequences
${\bf y}=(y_n)_{n\in \mathbb{Z}}\in \mathcal{F}$ and
\boldmath{$\xi$}\unboldmath$=(\xi_n)_{n\in
\mathbb{Z}}\in {\ker} D^*$. If $(\xi_n)_{n\in \mathbb{Z}}\in
\ker D^*$ then $\xi_n\in X_{*,n}$ for all $n\in \mathbb{Z}$ by the
definition of $X_{*,n}$. If $(y_n)_{n\in \mathbb{Z}}\in
\mathcal{F}$ then $y_n\in \Yn$ by the definition of
$\mathcal{F}$, and the claim is proved.

Next, fix ${\bf y}=(y_k)_{k\in\mathbb{Z}}\in \mathcal{F}\subset \im D$ 
and find
an ${\bf x}=(x_k)_{k\in \mathbb{Z}}\in \ell_p (\mathbb{Z};X)$,
resp., ${\bf x}\in c_0(\mathbb{Z};X)$, such that
$D{\bf x}={\bf y}$ or, in other words, such that for each $n\in \mathbb{Z}$
and all $k\in \mathbb N$ the following identity holds:
\begin{equation*}\begin{split}
x_n&=U(n,n-1)x_{n-1}+y_n=U(n,n-1)[U(n-1,n-2)x_{n-2}+y_{n-1}]+y_n\\
&=\dotsb
=U(n,n-k)x_{n-k}+\sum^{k-1}_{j=0}U(n,n-j)y_j.
\end{split}\end{equation*}
To prove the surjectivity of $D|_{\mathcal{F}}$ on
$\mathcal{F}$, we need to show that $x_n\in \Yn$ for
each $n\in\mathbb{Z}$. Fix $\xi \in X_{n,*}$ and pick a sequence
$(\xi_k)_{k\in \mathbb{Z}}\in \ker D^*$ such that $\xi=\xi_n$. By
\eqref{2.8} we have $U(n,n-k)^*\xi_n=\xi_{n-k}$. Since
$(y_k)_{k\in \mathbb{Z}}\in \mathcal{F}$, by Lemma~\ref{L4.1}(iv),
we have $U(n,n-j)y_j\in \Yn$ and $\langle U(n,n-j)y_j,\xi_n\rangle
=0$. Then
\begin{equation*}\begin{split} \langle x_n,\xi_n\rangle &=\langle
x_{n-k},U(n,n-k)^*\xi_n\rangle +\left\langle
\sum^{k-1}_{j=0}U(n,n-j)y_j,\xi_n\right\rangle\\
&=\langle x_{n-k},\xi_{n-k}\rangle \to 0\text{ as } k\to
\infty\end{split}\end{equation*} since $\| x_{n-k}\|\to 0$ 
 as $k\to \infty$ for the $\ell_p$--, resp.,
$c_0$--sequence ${\bf x}$.  Thus, $\langle x_n,\xi\rangle=0$
as claimed.
\hfill$\Box$\end{pf}

Recall that $X'_0$ is a direct complement of $X_0$ in 
$X_{0,*}^{\,\,\,\, \bot}$, see
\eqref{6.1}. Define the following closed subspace $\mathcal{F}_0$
of $\mathcal{F}$, see \eqref{6.2}:
$$\mathcal{F}_0:=\{(x_n)_{n\in \mathbb{Z}}\in \mathcal{F}:x_0\in X'_0\}.$$
Let $D_0$ denote the restriction $D|_{\mathcal{F}}$ acting on
$\mathcal{F}$ with the domain $\dom D_0=\mathcal{F}_0$.

\begin{lem}\label{L7.1} Operator $D_0$ is invertible on $\mathcal{F}$,
that is, for each $(z_n)_{n\in \mathbb{Z}}\in \mathcal{F}$ there
exists a unique $(x_n)_{n\in \mathbb{Z}}\in \mathcal{F}_0$ such that
$D(x_n)_{n\in \mathbb{Z}}=(z_n)_{n \in \mathbb{Z}}$.\end{lem}

\begin{pf} By Lemma~\ref{L6.1}, for each
${\bf z}=(z_n)_{n\in\mathbb{Z}}\in\mathcal{F}$ 
there exists a sequence ${\bf y}=(y_n)_{n\in
\mathbb{Z}}\in \mathcal{F}$ such that $D{\bf y}={\bf z}$. By the definition of
$\mathcal{F}$ we have $y_n\in \Yn$. Using the decomposition
$\Y0=X_0\oplus X'_0$, represent $y_0=y+y'$, where $y\in X_0$ and $y'\in
X'_0$. According to the definition of $X_0$, there exists a
sequence $(w_n)_{n\in \mathbb{Z}}\in {\ker} D$ such that
$w_0=y$. Let $x_n=y_n-w_n$, $n\in \mathbb{Z}$. Since $y_n\in \Yn$
and $w_n\in X_n\subset \Yn$, see Lemma~\ref{L4.1}(v), we infer that
${\bf x}=(x_n)_{n\in \mathbb{Z}}\in \mathcal{F}$. But
$x_0=y_0-w_0=y_0-y=y'\in X'_0$. Thus ${\bf x}\in
\mathcal{F}_0$. Since $(w_n)_{n\in \mathbb{Z}}\in \ker D$,
we also have $D{\bf x}=D{\bf y}={\bf z}$. To prove uniqueness, assume
that ${\bf x}\in \mathcal{F}_0$ and ${\bf x}\in \ker D$. 
By the definition of $X_n$ we have
$x_n\in X_n$ for all $n\in \mathbb{Z}$. In particular, $x_0\in
X_0$. But $(x_n)_{n\in \mathbb{Z}}\in \mathcal{F}_0$ means that
$x_0\in X'_0$. Since $X_0\cap X'_0=\{0\}$, we have $x_0=0$. Since
${\bf x}\in {\ker} D$, by \eqref{2.7} we
conclude that $x_n=U(n,0)x_0=0$ for $n\geq 0$. Also by \eqref{2.7},
we note that $0=x_0=U(0,n)x_n$ for $n<0$. By Lemma~\ref{L4.1}(ii),
$U(0,n)|_{X_n}:X_n\to X_0$, $n<0$, is invertible, and thus
$x_n\in X_n$ implies $x_n=0$ for $n<0$.\hfill$\Box$\end{pf}

\section{Punctured Dichotomies}\label{PD}
{\bf Dichotomy for {\boldmath$U(n,m)$\unboldmath}.}
We will now use the invertibility of $D_0$ on $\mathcal{F}$ to show that
the family of the restrictions
$U(n,m)|_{\Ym}:\Ym\to \Yn$ has a certain exponentially dichotomic
behavior on $\mathbb{Z}$ (a dichotomy on $\mathbb{Z}$ ``punctured''
at $m=0$). Recall that in this section $D$ is assumed to be Fredholm.

\begin{prop}\label{P9.1} There exist a family $\{P_n\}_{n\in
\mathbb{Z}}$ of projections defined on $\Yn$ such that $\sup_{n
\in \mathbb{Z}}\| P_n\|<\infty$, and constants $M\geq 1$ and $\alpha >0$
such that:
\begin{enumerate}\item[(i)] If $n\geq m>0$ or if
$ 0\geq n\geq m,$ then
\begin{equation}\label{9a.2}
P_n U(n,m)x=U(n_,m)P_mx \, \text{ for all } \, x\in \Ym.
\end{equation}
For the restriction $U(n,m)|_{\im P_m}:\im P_m\to
\im P_n$ we have:
\begin{equation}\label{9a.1} \| U(n,m)|_{\im P_m}\| \leq Me^{-\alpha
(n-m)};\end{equation}
\item[(ii)] If $n>0 \geq m$ and $x\in \Ym$, 
then $U(n,m)P_m x=P_n U(n,0)y'_0$, where $y'_0\in X'_0$ is the
component of $y=U(0,m)x$ in the representation $y=y_0+y'_0$,
 $y_0\in X_0$, corresponding to the direct sum decomposition
$\Y0=X_0\oplus X'_0$. Here, $X'_0$ is any direct complement of 
$X_0$ in $\Y0$;
\item[(iii)] If $n\geq m>0$ or if $0\geq n\geq m$ then the restriction
$U(n,m)|_{\ker P_m}:{\ker} P_m\to {\ker} P_n$ is an
invertible operator, and
$$
\| \left(U(n,m)|_{{\ker} P_m} \right)^{-1}\| \leq
Me^{-\alpha (n-m)};
$$
\item[(iv)] If $n>0\geq m$ then the {\em reduced} node operator $N(n,m)$ defined
as
$$
N(n,m):=(I-P_n)U (n,m)|_{{\ker} P_m}:{\ker} P_m\to 
\ker P_n
$$
is surjective with ${\ker} N(n,m)=X_m$.\end{enumerate}
\end{prop}

\begin{pf} Define on $\mathcal{F}$ a closed linear operator $T$ with
the domain $\dom T=\mathcal{F}_0$ by the rule $T:(x_n)_{n\in
\mathbb{Z}}\mapsto (U(n,n-1)x_{n-1})_{n\in \mathbb{Z}}$, such that
$D_0=I-T$. Note that although the domain of $T$ is {\em not} dense
in $\mathcal{F}$ (unless $X_0=\{0\}$), all standard facts from the spectral theory of
closed linear operators are still valid for $T$
(see~\cite[Ch. VII,\S9]{DS}). In particular, we can use the
spectrum, 
the resolvent set, and the
resolvent of $T$, that is, the operator $(\lambda I-T)^{-1}$,
bounded on $\mathcal{F}$, for $\lambda \in \rho(T)$.

For each $\lambda \in \mathbb{T}$, let
$V(\lambda)$ denote the isometry on $\mathcal{F}$ defined by the
rule $V(\lambda):(x_n)_{n\in \mathbb{Z}}\mapsto (\lambda^n
x_n)_{n\in \mathbb{Z}}$. Then
\begin{equation}\label{10.1} V(\lambda^{-1})TV(\lambda
)=\lambda^{-1}T,\quad |\lambda |=1.\end{equation} Thus, $\sigma
(T)=\mathbb{T} \cdot \sigma (T)$, that
is, $\sigma(T)$ is rotationally invariant. Since $1\in \rho(T)$ by
Lemma~\ref{L7.1}, we conclude that $\sigma (T)\cap \mathbb {T}=\emptyset$. Consider the Riesz
projection $\mathcal{P}=(2\pi i)^{-1}\int_{|\lambda |=1}
(\lambda -T)^{-1}d\lambda$ for $T$ on $\mathcal{F}$ that corresponds
to the  part of $\sigma (T)$ inside the unit disc:
\begin{equation}\label{10.2}
 \sigma (T|_{\im \mathcal{P}})=\sigma (T)\cap
\{ \lambda \in \mathbb{C}:|\lambda |<1\}.
\end{equation}
We stress that
$\mathcal{P}$ is a bounded operator on $\mathcal{F}$ and
$\im \mathcal{P}\subset \mathcal{F}_0$ since
$(\lambda-T)^{-1}(x_n)_{n\in \mathbb{Z}}\in \dom T=\mathcal{F}_0$ for
each $(x_n)_{n\in \mathbb{Z}}\in \mathcal{F}$ and $\lambda \in \mathbb {T}$. In
addition, the operator $T\mathcal{P}$ is defined on all of
$\mathcal{F}$ and is bounded, while the operator $\mathcal{P}T$ is
defined only on $\mathcal{F}_0;$ however, $T\mathcal{P}\supset
\mathcal{P}T$, that is,
\begin{equation}\label{10.3}
T\mathcal{P}(x_n)_{n\in \mathbb{Z}}=\mathcal{P}T(x_n)_{n\in
\mathbb{Z}} \, \text{ for all } \, (x_n)_{n\in \mathbb{Z}}\in
\mathcal{F}_0.
\end{equation}
Also, by \eqref{10.2}, $\sprad (T|_{\im \mathcal{P}})<1$. The
restriction $T|_{\ker \mathcal{P}}$ is an operator on $\ker
\mathcal{P}$ with the domain $\dom T|_{\ker
\mathcal{P}}={\ker} \mathcal{P}\cap \mathcal{F}_0$ and with
the spectrum $\sigma (T|_{{\ker} \mathcal{P}})=\sigma
(T)\cap \{ \lambda \in \mathbb{C}:|\lambda |>1\}$. In particular,
$T|_{{\ker} \mathcal{P}}$ is invertible in $\ker\mathcal{P}$ and $\sprad
((T|_{{\ker} \mathcal{P}})^{-1})<1$. Fix any positive
$\alpha $ strictly smaller than $$-\ln \max \{ \sprad
(T|_{\im \mathcal{P}}), \sprad ((T|_{{\ker
} \mathcal{P}})^{-1})\}.$$ Thus, there is a constant 
$M\geq 1$ such that:
\begin{equation}\label{11.1}
\| (T|_{\im \mathcal{P}})^k\| \leq Me^{-\alpha k}\text{ and } \|
(T|_{{\ker} \mathcal{P}})^{-k}\| \leq Me^{-\alpha k},\quad k\in
\mathbb{Z}_+.
\end{equation}
Next, we claim that there exists a family $\{ P_n\}_{n\in
\mathbb{Z}}$ of projections on $\Yn$ such that
$\sup_{n\in\mathbb{Z}}\| P_n\|<\infty$ and
$\mathcal{P}=\diag_{n\in \mathbb{Z}}[P_n]$, that is, for each
$(x_n)_{n\in \mathbb{Z}}\in \mathcal{F}$ we have
$\mathcal{P}(x_n)_{n\in \mathbb{Z}}=(P_nx_n)_{n\in \mathbb{Z}}$.
Indeed, \eqref{10.1} and the integral formula for $\mathcal{P}$
imply $ V(\lambda^{-1})\mathcal{P}V(\lambda)=\mathcal{P}$
for all $\lambda \in\mathbb T$.
Since $\mathcal{P}$ commutes with the family $\{ V(\lambda ):
|\lambda |=1 \}$, by \cite[Lem.~3]{Ba} we conclude that
$\mathcal{P}$ is a diagonal operator, that is,
$\mathcal{P}=\diag_{n\in \mathbb{Z}}[P_n]$. The operators $P_n$
here are defined as follows: fix an $x\in \Yn$ and 
define $P_n x$ as the $n$-th element in the
sequence $\mathcal{P}(x\otimes{\bf e}_n)$. Note that
$\sup_{n\in \mathbb{Z}}\| P_n\|=\|\mathcal{P}\| <\infty$, and the
claim is proved.

Fix $m\in \mathbb{Z}$, take any $x\in \Ym$, 
and let ${\bf x}=x\otimes{\bf e}_m$. 
Note that ${\bf x}\in \mathcal{F}_0$ provided either
$m\neq 0$ or $m=0$ and $x\in X'_0$. If ${\bf x}\in \mathcal{F}_0$
then \eqref{10.3} implies:
$$T \mathcal {P}{\bf x}=U(m+1,m)P_m x\otimes {\bf e}_{m+1}
=\mathcal{P}T{\bf x}=P_{m+1}U(m+1,m)x\oplus{\bf e}_{m+1}.$$
Thus if $m\neq 0$, or if $m=0$ and $x \in X'_0$, then
$U(m+1,m)P_mx=P_{m+1}U(m+1,m)x$.
Recall that if $n>m$ then $U(n,m)=U(n,m+1)U(m+1,m)$. Using this we derive
\eqref{9a.2}. For ${\bf x}=x\otimes {\bf e}_m$ we note that
$T^j{\bf x}=U(m+j,m)x\otimes {\bf e}_{m+j}\in \mathcal{F}_0$ for
$j=0,1,\dotsc ,n-m$ provided either $n\ge m>0$, or $0>n\ge m$, 
or $n=0\ge m$ and $U(0,m)x\in X_0'$. Then the 
first inequality in \eqref{11.1} implies \eqref{9a.1}, and (i) in
Proposition \ref{P9.1} is proved.
\begin{lem}\label{XnNR} The following inclusions hold:
\begin{equation}\label{12.3}
 X_n\subset {\ker} P_n \, \text{ for }\, n\leq 0 \, \text{
and } \, X_n\subset \im P_n \, \text{ for } \, n>0.
\end{equation}
\end{lem}
\begin{pf}
We present the proof for the $\ell_p$--case, the $c_0$--case is similar.
By \eqref{3.1} and \eqref{2.7}, if $x\in X_n$ then
there is a sequence $(x_n)_{n\in \mathbb{Z}}\in
\ell_p(\mathbb{Z};X)$ such that $x=x_n$ and $x_n=U(n,m)x_m$ for
all $n\geq m$ in $\mathbb{Z}$. Note that $\mathcal P (x_n)_{n\in
\mathbb{Z}}=(P_nx_n)_{n\in \mathbb{Z}}\in \im \mathcal P\subset
\mathcal{F}_0$ and thus by \eqref{10.3} we have 
$T^k\mathcal{P}(x_n)_{n\in \mathbb{Z}}\in \im \mathcal{P}\subset
\mathcal{F}_0$ for all
$k \in \mathbb N$. If $(y_n)_{n\in \mathbb{Z}}=T^k(P_nx_n)_{n\in
\mathbb{Z}}$, where $y_n=y_n(k)$,
 then $y_n=U(n,n-k)P_{n-k}x_{n-k}$. Using
\eqref{9a.2}, we have that if $n-k>0$ or $0\geq n$ then
$y_n=U(n,n-k)P_{n-k} x_{n-k}=P_nU(n,n-k)x_{n-k}$. But $(x_n)_{n\in
\mathbb{Z}}\in {\ker}\, D$ and thus $U(n,n-k)x_{n-k}=x_n$. So, finally,
\begin{equation}\label{12.4}
y_n=P_nx_n\text{ for } n>k\text{ or } 0\geq n.\end{equation} By
the first inequality in \eqref{11.1} we know that
$$\lim_{k\to
\infty}\| (y_n)_{n\in \mathbb{Z}}\|_{\ell_p}=\lim_{k\to \infty} \|
T^k(P_nx_n)_{n\in\mathbb{Z}}\|_{\ell_p}=0.$$ But using \eqref{12.4} we have:
$$\| (y_n)_{n\in \mathbb{Z}}\|^p_{\ell_p}=\sum_{n\in \mathbb{Z}}\|
y_n\|^p \geq \sum_{n\leq 0}\| y_n\|^p=\sum_{n\leq 0}\|
P_nx_n\|^p.$$ So, $P_nx_n=0$, that is, $X_n\subset {\ker}\, P_n$ for
$n\leq 0$.

To prove the  second inclusion in \eqref{12.3}, note that
$((I-P_n)x_n)_{n\in \mathbb{Z}}\in {\ker} \, \mathcal{P}$. Since
$T|_{{\ker} \mathcal{P}}$ is invertible on ${\ker
} \, \mathcal{P}$ and the second inequality in \eqref{11.1} holds, for
each $k \in \mathbb N$ there exists a sequence $(y_n)_{n\in
\mathbb{Z}}\in \mathcal{F}_0\cap {\ker} \, \mathcal P$, 
where $y_n=y_n(k)$, such that
$T^k(y_n)_{n\in \mathbb{Z}}=((I-P_n)x_n)_{n\in \mathbb{Z}}$ and
\begin{equation}\label{13.1} \lim_{k\to \infty} \| (y_n)_{n\in
\mathbb{Z}}\|_{\ell_p}=\lim_{k\to \infty} \|(T|_{{\ker
} \, \mathcal{P}})^{-k}((I-P_n)x_n)_{n\in
\mathbb{Z}}\|_{\ell_p}=0.\end{equation} Using the equality
$x_n=U(n,m)x_m$ and \eqref{9a.2}, we find that if $n-k>0$ or if
$0\ge n$ then the $n$-th element of the sequence $T^k(y_n)_{n\in
\mathbb{Z}}=((I-P_n)x_n)_{n\in \mathbb{Z}}$ is equal to
\begin{equation*}
\begin{split}U(n,n-k)y_{n-k}&=(I-P_n)x_n=(I-P_n)U(n,n-k)x_{n-k}
\\
&=U(n,n-k)(I-P_{n-k})x_{n-k}.\end{split}\end{equation*}
In other words, $y_{n-k}-(I-P_{n-k})x_{n-k}\in {
\ker} \, U(n,n-k)$. We claim that, in fact, this implies that
\begin{equation}\label{13.2} y_{n-k}-(I-P_{n-k})x_{n-k}=0\text{ provided
} n>k.\end{equation}
As soon as the claim is proved, we write:
\begin{equation*}\begin{split} \| (y_n)_{n\in
\mathbb{Z}}\|^p_{\ell_p}&=\| (y_{n-k})_{n\in
\mathbb{Z}}\|^p_{\ell_p}\geq \sum_{n>k}\|y_{n-k}\|^p\\&=
\sum_{n>k}\| (I-P_{n-k})x_{n-k}\|^p=\sum_{n>0}\| (I-P_n)x_n\|^p.
\end{split}\end{equation*}
Now \eqref{13.1} implies $(I-P_n)x_n=0$, that is, $X_n\subset {
\ker } \, P_n$ for $n>0$. It remains to prove the claim \eqref{13.2}.
Recall that $(y_n)_{n\in \mathbb{Z}}\in {\ker} \, \mathcal{P}$
and thus $y_{n-k}-(I-P_{n-k})x_{n-k}\in \, {\ker} \, P_{n-k}$ for
$n>k$. So, it suffices to check that ${\ker} \, U(n+k,n)\cap {
\ker} \, P_n =\{0\}$ for all $n>0$ and any $k>0$. If $n>0$ and $x\in
{\ker} \, U(n+k;n)\cap {\ker } \, P_n$ then the sequence ${\bf
x}=x\otimes {\bf e}_n$ belongs to $\ker \, \mathcal{P} \cap
\mathcal{F}_0$. Note that for $j \in \mathbb N$ we have $T^j{\bf
x}=U(n+j,n)x\otimes {\bf e}_{n+j}$. Thus, $T^k{\bf x}=0$ since
$U(n+k,n)x=0$. Now the second inequality in \eqref{11.1} implies
that
$0=\|T^k{\bf x}\|_{\ell_p}\geq M^{-1}e^{\alpha k}\|{\bf
x}\|_{\ell_p}=M^{-1}e^{\alpha k}\|x\|$. Thus,
claim \eqref{13.2} is proved, and the proof of the inclusions
\eqref{12.3} and Lemma \ref{XnNR} is finished.\hfill$\Box$\end{pf}

To prove (ii) in Proposition \ref{P9.1}, we first consider
$n=1$ and $m=0$. We can now apply \eqref{10.3}  for
$(x_n)_{n \in \mathbb Z}=x\otimes{\bf e}_0$ only when
 $x\in X'_0$, and obtain
$U(1,0)P_0x=P_1U(1,0)x$ provided $x\in X'_0$. 
This implies that if $n > m=0$ then
\begin{equation}\label{12.2}
 U(n,0)P_0x=P_nU(n,0)x \, \text{ for all } \, x\in X'_0.\end{equation}
Next, for $n>0\ge m$,
fix $x\in \Ym$ and denote $y=U(0,m)x$.
Using the equality $U(0,m)P_mx=P_0U(0,m)x$ from \eqref{9a.2}, we have
$U(n,m)P_mx=U(n,0)U(0,m)P_mx=U(n,0)P_0y$. 
Represent $y=y_0+y'_0$,
where $y_0\in X_0$ and $y'_0\in X'_0$, and recall that $P_0y_0=0$
by \eqref{12.3} in Lemma \ref{XnNR}. Then, using equation~\eqref{12.2}, we conclude:
$U(n,m)P_mx=U(n,0)P_0(y_0+y'_0)
=U(n,0)P_0 y'_0=P_n U(n,0)y'_0$,
and (ii) in Proposition \ref{P9.1} is proved.

To prove (iii) in Proposition \ref{P9.1},
remark that by the second inequality in \eqref{11.1} we have
the inequality $\|(T|_{\ker \mathcal{P}})^{-k}(x_n)_{n\in
\mathbb{Z}}\|_{\mathcal{F}} \leq
Me^{-\alpha k}\| (x_n)_{n\in \mathbb{Z}}\|_{\mathcal{F}}$. As soon as
$T^j(x_n)_{n\in \mathbb{Z}}\in \ker \mathcal{P}\cap \mathcal{F}_0$
for $j=0,1,\ldots k-1$, we then have $\|T^k(x_n)_{n\in
\mathbb{Z}}\|_{\mathcal{F}}\geq M^{-1}e^{\alpha k}
\| (x_n)_{n\in \mathbb{Z}}\|_{\mathcal{F}}$.
In particular,
$T^j(x\otimes {\bf
e}_m)=U(n+j,m)x \otimes {\bf e}_{m+j}\in \mathcal{F}_0$ if and
only if either $m>0$, or $m+j<0$, or $m=-j$ and $U(0,-j)x\in
X'_0$. This implies that $\| U(m+k,m)x\|\geq M^{-1}e^{\alpha
k}\|x\|$ provided one of the following three possibilities hold:
(a) $m>0$, \, $k \in \mathbb Z_+$, \, $x\in {\ker} \, P_m$;
(b) $m<0$, \, $k=0,1,\ldots ,-m$, \, $x\in {\ker} \, P_m$;
(c) $m=0$, \, $k \in \mathbb Z_+$, $x\in X'_0\cap {\ker} \, P_0$.
This proves (iii).

To prove (iv) in Proposition \ref{P9.1}, 
we first consider the reduced node operator
$N(1,0)=(I-P_1)U(1,0)|_{{\ker} \, P_0} : {\ker } \, P_0 \to {
\ker } \, P_1$. Note that
$ {\ker}\, N(1,0)=\{x\in {\ker} \, P_0 : U(1,0)x \in \im \, P_1\}$.
We claim that $X_0=\ker \, N(1,0)$. Indeed,
$U(1,0)(X_0)=X_1\subset \im \, P_1$ by Lemma~\ref{L4.1}(ii) and
\eqref{12.3}, which implies $X_0\subset {\ker} \, N(1,0)$. To
prove the inverse inclusion, assume that $x\in {\ker} \, P_0$ and
$U(1,0)x\in \im \, P_1$. Using $\Y0=X_0\oplus X'_0$, decompose
$x=x_0+x'_0$. Then $U(1,0)x'_0=U(1,0)x-U(1,0)x_0\in \im \, P_1$ since
$U(1,0)x\in \im P_1$ by assumption and $U(1,0)x_0\in X_1\subset
\im P_1$ by Lemma \ref{L4.1} (ii) and \eqref{12.3}. 
Also, $x'_0\in {\ker} \, P_0\cap X'_0$
since $x'_0=x-x_0$ and $x\in {\ker} \, P_0$ by assumption, and
$x_0\in X_0\subset {\ker} \, P_0$ by \eqref{12.3}. Therefore,
$x'_0\otimes {\bf e}_0\in {\ker} \mathcal{P} \cap \mathcal{F}_0$ 
and, using \eqref{11.1}, we obtain for $k\in\mathbb{N}$:
\begin{equation*}\begin{split} \| U(k,1)U(1,0)x'_0\| &= \| U(k,0)x'_0\|
= \| U(k,0)x'_0\otimes {\bf e}_k\|_{\ell_p}\\
&= \| T^k (x'_0\otimes {\bf e}_0)\|_{\ell_p}\geq M^{-1}e^{\alpha
k}\| x'_0\otimes {\bf e}_0\|_{\ell_p}=M^{-1}e^{\alpha
k}\|x'_0\|.\end{split}\end{equation*} But then \eqref{9a.1} for
$U(1,0)x'_0\in \im P_1$  implies $\| x'_0\|=0$ and thus $x=x_0$
proving $\ker N(1,0)\subset X_0$.

Next, we show that for each $y\in  \ker \, P_1$ there is an $x\in {
\ker} \, P_0$ such that $(I-P_1)U(1,0)x=y$. Take $y\otimes {\bf
e}_1\in {\ker} \mathcal{P}$ and find $(x_n)_{n\in
\mathbb{Z}}\in {\ker} \mathcal{P}\cap \mathcal{F}_0$ so that
$T(x_n)_{n\in \mathbb{Z}}=y\otimes {\bf e}_1$. In particular,
$U(1,0)x_0=y$ for $x_0\in {\ker} \, P_0\cap X'_0$. Then $y=(I-P_1)y
=(I-P_1)U(1,0)x_0$, and $N(1,0)$ is
surjective from ${\ker} \, P_0$ to ${\ker} \, P_1$ with ${
\ker} \, N(1,0)=X_0$.

To finish the proof of (iv) in Proposition \ref{P9.1}
for any $n>0\geq m$, we remark that
$U(n,m)=U(n,1)U(1,0)U(0,m)$ and \eqref{9a.2} imply:
$(I-P_n)U(n,m)(I-P_m)=[(I-P_n)U (n,1)(I-P_1)]N(1,0)[(I-P_0)U(0,m)(I-P_m)]$.
Operators in brackets are invertible by (iii), and the general case
$n>0\geq m$ in (iv) follows from the case $n=1$ and $m=0$ proved above.
\hfill$\Box$\end{pf}

{\bf Dichotomy for {\boldmath$U(n,m)^*$\unboldmath}.}
In addition to Proposition \ref{P9.1}, 
for the proof of Theorem~\ref{FREDSA} we will need to 
consider the following dual objects.
For $k\geq \ell$ in $\mathbb{Z}$ define an 
exponentially bounded evolution family $\{ U_*(k,\ell)\}_{k\geq \ell}$ 
on $X^*$ by  $U_*(k,\ell)=U(-\ell, -k)^*$. Let
$D_*:(\xi_k)_{k\in \mathbb{Z}}\mapsto 
(\xi_k-U_*(k,k-1)\xi_{k-1})_{k\in \mathbb{Z}}$
denote the corresponding difference operator. 
Also, consider an operator, $D_\sharp$,
defined by the rule $D_\sharp:(\xi_n)_{n\in \mathbb{Z}}\mapsto
(\xi_n-U(n+1,n)^*\xi_{n+1})_{n\in \mathbb{Z}}$ on the following spaces:
If $D$ is acting on $\ell_p$, $p\in(1,\infty)$,
then $D_\sharp$ is considered on $\ell_{q,*}$, $q\in(1,\infty)$,
$p^{-1}+q^{-1}=1$, and then $D_\sharp=D^*$, the adjoint operator of $D$.
If $D$ is acting on $\ell_1$, then $D_\sharp$ is considered on $c_{0,*}$, and
then $(D_\sharp)^*=D$. If $D$ is acting on $c_0$, then $D_\sharp$ is
considered on $\ell_{1,*}$, and then $D_\sharp=D^*$. 
If $j:(\xi _k)_{k\in \mathbb{Z}}
\mapsto (\xi _{-k})_{k\in \mathbb{Z}}$,
and the operator $D_*$ is considered on the same sequence space
as $D_\sharp$, then $D_*=jD_\sharp j^{-1}$. 
Since $D$ is Fredholm if and only if $D^*$ is Fredholm, 
we infer that $D_\sharp$ is Fredholm, 
and therefore $D_*$ is Fredholm. Moreover,
$\ind D_* =\ind D$. 
Apply \eqref{2.7}-\eqref{2.8} for $D_*$ and 
$\{ U_*(k,\ell)\}_{k\geq \ell}$, and remark that 
$U_*(k,\ell)^*=U(-\ell,-k)$ acts on $X$ by the reflexivity assumption.
Then, for sequences $(\xi_k)_{k\in\mathbb{Z}}$ and $(z_k)_{k\in\mathbb{Z}}$
from the corresponding sequence spaces, we infer:
\begin{align}\label{SS1.1}\ker \, D_*&=\{(\xi _k)_{k\in \mathbb{Z}}:
\xi_k=U_*(k,\ell)\xi_{\ell}, \quad k\geq \ell\},\\
\label{SS1.2}\ker  (D_*)^*&=\{ (z_k)_{k\in \mathbb{Z}}:
z_\ell=U_*(k,\ell)z_k, \quad k\geq \ell\}.\end{align}
For $k\in \mathbb{Z}$ introduce subspaces $Z_{k,*}\subset X^*$, resp. $Z_k\subset X$, resp. $Z_k^{\bot}\subset X^*$, for $\{ U_*(k,\ell)\}_{k\geq \ell}$ that are
analogous to  the subspaces $X_n\subset X$, 
resp. $X_{n,*}\subset X^*$, resp. $\Yn\subset X$, 
for $\{U(n,m)\}_{n\geq m}$, defined in \eqref{3.1} and \eqref{3.2}:
\begin{align} Z_{k,*}&=\{ \xi \in X^*:\text{ there exists } (\xi_\ell )_{\ell 
\in \mathbb{Z}}\in \ker \, D_*\text{ so that } \xi=\xi_k\},\label{S1.1}\\
Z_k&=\{ z\in X: \text{ there exists } (z_{\ell})_{\ell \in \mathbb{Z}}\in 
\ker (D_*)^*\text{ so that } z=z_k\}\label{S1.2}.\end{align}
\begin{lem}\label{LS2.1} For each $k\in \mathbb{Z}$ we have
 $Z_k=X_{-k}$ and $Z_{k,*}=X_{-k,*}$.\end{lem}
\begin{pf} By formulas \eqref{SS1.2} and \eqref{S1.2}, 
$z\in Z_k$ if and only if $z=z_k$ for a sequence 
$(z_\ell)_{\ell \in \mathbb{Z}}$ such that
$z_\ell =U_*(k,\ell)^*z_k=(U(-\ell -k)^*)^*z_k=U(-\ell ,-k)z_k$
for all $k\geq \ell$.
By formulas \eqref{2.7} and \eqref{3.1}, $x\in X_m$ if and only if 
$x=x_m$ for a sequence $(x_n)_{n\in \mathbb{Z}}$ such that
$x_n=U(n,m)x_m$ for all $n\geq m$. Setting $z_{-n}=x_n$, $n\in \mathbb{Z}$, thus proves 
$Z_k=X_{-k}$. The proof of $Z_{k,*}=X_{-k,*}$ is similar.\hfill$\Box$\end{pf}
Apply Proposition~\ref{P9.1} to the 
evolution family $\{U_*(k,\ell)\}_{k\geq\ell}$.
This proposition gives the following assertions:
a dichotomy for the restriction
$\{U_*(k,\ell)|_{Z^\bot_\ell}\}_{k\geq\ell}$ for $k\ge\ell>0$ and
$0\ge k\ge\ell$, an analogue of Lemma \ref{XnNR}, and the surjectivity of
the reduced node operator that corresponds to this restriction.
Using Lemma~\ref{LS2.1}, and setting $n=-\ell$ and $m=-k$ for 
$n\geq m$ in $\mathbb{Z}$, 
we now recast these assertions for the family $\{U(n,m)^*|_{X^\bot_n}\}_{n\ge
m}$ as follows (cf. Proposition \ref{P9.1} and Lemma \ref{XnNR}).
\begin{prop}\label{PS3.1} There exist 
 a family $\{P_{n,*}\}_{n\in \mathbb{Z}}$ of projections 
 defined on $X^\bot_n$ 
such that $\sup_{n\in \mathbb{Z}}\| P_{n,*}\|<\infty$,
and constants $M\geq 1$ and $\alpha >0$ such that:
\begin{enumerate}
\item[(i)] If $n\ge m\ge 0$ or if $0>n\ge m$ then 
\begin{equation}
\label{S3.3} P_{m,*} U(n,m)^*\xi =U(n,m)^*P_{n,*}\xi \text{ for all } 
\xi \in X^\bot_n.\end{equation}
For the restriction 
$U(n,m)^*|_{\im P_{n,*}}:\im P_{n,*}\to \im P_{m,*}$ we have:
\begin{equation}\label{S3.4}
 \| U(n,m)^*|_{\im P_{n,*}}\|\leq Me^{-\alpha (n-m)};\end{equation}
\item[(ii)] If $n\geq 0>m$ and $\xi \in X^\bot_n$, then
\begin{equation}\label{SN12.3} U(n,m)^*P_{n,*}\xi=
P_{m,*}U(0,m)^*\zeta '_0,\end{equation}
where $\zeta'_0\in X'_{0,*}$ is the component of 
$\zeta =U(n,0)^*\xi$ in the representation 
$\zeta=\zeta_0+\zeta'_0$, $\zeta_0\in X_{0,*}$, 
corresponding to the direct sum decomposition
$X^\bot_0=X_{0,*}\oplus X'_{0,*}$. Here, $X'_{0,*}$ is any direct
complement of $X_{0,*}$ in $X_0^\bot$;
\item[(iii)] If $n\ge m\ge 0$ or if $0>n\ge m$ then
the restriction 
$U(n,m)^*|_{\ker \, P_{n,*}}:\ker P_{n,*}\to \ker \, P_{m,*}$ 
is an invertible operator, and
\begin{equation}\label{S3.5} 
\| (U(n,m)^*|_{\ker \, P_{n,*}})^{-1}\| \leq Me^{-\alpha
(n-m)};\end{equation}
\item[(iv)] If $n\geq 0>m$ then the reduced node operator
$N_*(n,m)$ defined as
\begin{equation}\label{SN12.5} N_*(n,m)=(I-P_{m,*})U(n,m)^*|_{\ker \, P_{n,*}}:
\ker \, P_{n,*}\to \ker \, P_{m,*}\end{equation}
is surjective with $\ker N_*(n,m)=X_{n,*}$.
\item[(v)] The following inclusions hold:
\begin{equation}\label{S3.6}
X_{n,*}\subset \ker P_{n,*}\text{ for } 
n\geq 0\text{ and } X_{n,*}\subset \im \, P_{n,*}\text{ for } 
n<0.\end{equation}
\end{enumerate}
\end{prop}
{\bf Invariant direct complements.} Recall the direct sum decomposition  
$X=\Yn\oplus Y_n$, see \eqref{6.1}. It allows us to identify
\begin{equation}\label{S4.1} (Y_n)^*=(\Yn)^\bot=X_{n,*},\quad n\in \mathbb{Z}.\end{equation}
Recall that $\dim X_{n,*}<\infty$ by Lemma~\ref{L4.1}(i) and 
thus $X_{n,*}$ has a direct complement in $X^*$. Let $Q_{n,*}$ be a 
bounded projection on $X^*$ such that $\im
Q_{n,*}=X_{n,*}$. By Lemma~\ref{L4.1}(iii) 
we have $U(n,m)^*(X_{n,*})\subseteq X_{m,*}, n \ge m,$ or
\begin{equation} U(n,m)^*Q_{n,*}=Q_{m,*}U(n,m)^*Q_{n,*}.\label{S4.2}\end{equation}
Note that $Y_n$ is an arbitrary direct complement of the finitely codimensional subspace $\Yn$ in $X$, 
and, generally, $U(n,m)(Y_m)\nsubseteq Y_n$. Using
representation \eqref{1B.1} with $P=Q_{m,*}$ and $Q=Q_{n,*}$
for $A=U(n,m)^*$ in the decompositions $X^*=\im Q_{m,*}\oplus 
\ker Q_{m,*}$ and $X^*=\im Q_{n,*}\oplus 
\ker Q_{n,*}$, we will identify the restriction $U(n,m)^*|_{X_{n,*}}$
and the operator $U(n,m)^*Q_{n,*}:X_{n,*}\to X_{m,*}$. This is a finite dimensional and, 
by Lemma~\ref{L4.1}(iii), invertible operator. 
By \eqref{S4.1} and \eqref{S4.2},
$(U(n,m)^*Q_{n,*})^*=Q^*_{n,*}U(n,m)Q^*_{m,*}:Y_m\to Y_n$.

If $n\geq 0$ then $X_{n,*}\subset \ker P_{n,*}$ by \eqref{S3.6} and thus \eqref{S3.5} implies
\begin{equation}\|U(n,m)^*\xi \| \geq M^{-1}e^{\alpha (n-m)}\| \xi \|\text{ for all } \xi \in X_{n,*}.\label{S4.4}\end{equation}
Hence, $\| (U(n,m)^*Q_{n,*})^{-1}\| _{\mathcal{L}(X_{m,*},X_{n,*})}\leq Me^{-\alpha (n-m)}$. 
Passing to the adjoint in \eqref{S4.2}, and using \eqref{S4.1}, we
conclude that the operator 
\begin{equation}Q^*_{n,*}U(n,m)=Q^*_{n,*}U(n,m)Q^*_{m,*}:Y_m\to Y_n\label{S4.3}\end{equation}
is invertible, and
\begin{equation}\label{S5.1} 
\| (Q^*_{n,*}U(n,m)Q^*_{m,*})^{-1}\|_{\mathcal{L}(Y_n,Y_m)}
\leq Me^{-\alpha (n-m)},\quad n\geq m\geq 0,\end{equation}
for any direct complement $Y_n$ of $\Yn$ in $X$. 
Next, we will identify the direct complement of $\Yn$ in $X, n \ge 0,$ which is $U(n, m)$-invariant.
Fix any $Y_0$ such that $\Y0\oplus Y_0=X$. For each $n\geq 0$ define
$W_n:=\{U(n,0)y_0:y_0\in Y_0\}$.
\begin{lem}\label{SL5.1} 
For all $n\ge m\ge0$ in $\mathbb{Z}_+$ the following 
assertions hold:
\begin{enumerate}\item[(i)] the subspace $W_n$ is closed;
\item[(ii)] $\Yn\oplus W_n=X$;
\item[(iii)] $U(n,m)W_m\subseteq W_n, n\geq m\geq 0$;
\item[(iv)] the restriction $U(n,m)|_{W_m}:W_m\to W_n$ is invertible, and
\begin{equation}\label{S5.2} \| (U(n,m)|_{W_m})^{-1}\| \leq Me^{-\alpha (n-m)}.\end{equation}
\end{enumerate}\end{lem}
\begin{pf} {(i)} Inequality \eqref{S5.1} and \eqref{S4.3} for 
$n\geq m=0$ imply for all $y_0\in Y_0$:
\begin{equation}\label{S5.3}M^{-1}e^{\alpha n}
\| y_0\|\leq \|Q^*_{n,*}U(n,0)Q^*_{0,*}y_0\| =
\| Q^*_{n,*}U(n,0)y_0\| \leq \|Q^*_{n,*}\| \| U(n,0)y_0\|.\end{equation}
Thus, $\| U(n,0)y_0\| \geq c\| y_0\|$ for some $c>0$, and (i) holds.

{(ii)} To see $\Yn\cap W_n=\{0\}$, assume that 
$x=U(n,0)y_0\in \Yn$ for some $y_0\in Y_0$. Since $U(n,0)^*:X_{n,*}\to X_{0,*}$ is an isomorphism by
Lemma~\ref{L4.1}(iii), if $\xi_0\in X_{0,*}$ then $\xi_0=U(n,0)^*\xi_n$ 
for some $\xi_n\in X_{n,*}$. Since $x\in \Yn$, for each $\xi_0\in X_{0,*}$ we have:
$\langle y_0,\xi_0\rangle =
\langle y_0,U(n,0)^*\xi_n\rangle =
\langle U(n,0)y_0,\xi_n\rangle =\langle x,\xi_n\rangle=0$.
Thus, $y_0\in \Y0\cap Y_0$ and $y_0=0=x$.

To see $(W_n+\Yn)^\bot=W^\bot_n\cap X_{n,*}=\{0\}$, 
assume that $\xi_n\in W^\bot_n\cap X_{n,*}$. 
Then for each $y_0\in Y_0$ and $x=U(n,0)y_0\in W_n$ we have
$0=\langle \xi _n,x\rangle =\langle \xi_n,U(n,0)y_0\rangle 
=\langle U(n,0)^*\xi_n,y_0\rangle $.
Thus, $U(n,0)^*\xi_n\in (Y_0)^\bot$. On the other hand, 
$\xi_n\in X_{n,*}$ and Lemma~\ref{L4.1}(iii) imply 
$U(n,0)^*\xi_n\in X_{0,*}$. Thus
$U(n,0)^*\xi_n=0$ and $\xi_n=0$ by Lemma \ref{L4.1}(iii), 
which finishes the proof of (ii).

(iii) If $x=U(m,0)y_0\in W_m$ then $U(n,m)x=U(n,0)y_0\in W_n$.

(iv) By (ii), we have $(W_n)^*=(X^\bot_{n,*})^\bot =X_{n,*}$. 
By (iii), we are in the situation when $U(n,m)^*|_{X_{n,*}}:X_{n,*}\to
X_{m,*}$ is the adjoint of the operator $U(n,m)|_{W_m}:W_m\to W_n$. 
By \eqref{S4.4}, both (finite dimensional) operators are invertible, the norms of inverses
are equal, and thus \eqref{S4.4} implies \eqref{S5.2}.
\hfill$\Box$\end{pf}
We proceed further with a construction of the direct complement of $X^\bot_n, n \le 0,$
in $X^*$ which is $U(n,m)^*$-invariant.
 Consider  a direct 
sum decomposition $X^*=X^\bot_n\oplus Y_{n,*}$, $n \le 0$, 
where $Y_{n,*}$ is any direct
complement of the (finitely codimensional) subspace $X^\bot_n$ in $X^*$. 
We may identify
$(Y_{n,*})^*=(X^\bot_n)^\bot =X_n$.
Define $W_{n,*}=\{ U(0,n)^*\xi_0 :\xi_0 \in Y_{0,*}\}$, $n\leq 0$.

\begin{lem}\label{SL9.1}  
For all $m\le n\le 0$ in $\mathbb{Z}_-$ 
the following assertions hold:
\begin{enumerate}\item[(i)] The subspace $W_{n,*}$ is closed;
\item[(ii)] $X^\bot_n\oplus W_{n,*}=X^*$;
\item[(iii)] $U(n,m)^*W_{n,*}\subseteq W_{m,*}$;
\item[(iv)] the restriction $U(n,m)^*|_{W_{n,*}}:W_{n,*}\to W_{m,*}$ is invertible, and 
\begin{equation}\label{S9.2} \| (U(n,m)^*|_{W_{n,*}})^{-1}\| \leq Me^{-\alpha (n-m)}.\end{equation}\end{enumerate}\end{lem}

\begin{pf} The proof is parallel to the proof of Lemma~\ref{SL5.1}. Indeed, the inclusion $X_n\subset \ker P_n$, $n\leq 0$, in \eqref{12.3} and
Proposition~\ref{P9.1}(iii) imply that $U(0,n)|_{X_n}:X_n\to X_0$ is 
invertible with $\| (U(0,n)|_{X_n})^{-1}\|\leq Me^{\alpha n}$, $n\leq 0$. 
Using any bounded
projection $Q_n$ on $X$ with $\im Q_n=X_n$, we identify $U(0,n)|_{X_n}=
U(0,n)Q_n=Q_0U(0,n)Q_n$.
 Passing to the adjoint operator, cf. \eqref{S5.3}, we conclude that $\|
U(0,n)^*\xi \| \geq c\| \xi \|$ for all $\xi \in Y_{0,*}=(X_0)^*$. 
This gives { (i)}, and the proof of {(ii)--(iv)} is identical (dual) to the proof of
Lemma~\ref{SL5.1}.\hfill$\Box$\end{pf}
\section{Proof of Theorem \ref{FREDSA}}\label{FDOED}
\noindent {\bf PROOF OF THEOREM~\ref{FREDSA} FOR 
{\boldmath$n\ge0$\unboldmath}.}\quad 
First, consider $n>0$. By Proposition~\ref{P9.1} 
and Lemma~\ref{SL5.1}(ii) we have a direct sum decomposition
$X=\Yn\oplus W_n=\im P_n\oplus \ker P_n\oplus W_n$, $n>0$.
Let $P^+_n$ be a projection on $X$ with \begin{equation}\label{S7N.1}\im P^+_n=\im P_n\text{ and }
\ker P^+_n=\ker P_n\oplus W_n, \quad n>0.\end{equation} For
$n\geq m>0$, if
$x\in\im \, P^+_m$ then
$U(n,m)x\in\im \, P^+_n$ by \eqref{9a.2}. If $x=y+z\in \ker \, P^+_m$, 
where $y\in \ker \, P_m$, $z\in W_m$, then $U(n,m)x=U(n,m)y+U(n,m)z\in \ker \, P^+_n$ by \eqref{9a.2} and
Lemma~\ref{SL5.1}(iii). This gives
$U(n,m)P^+_m=P^+_nU(n,m)$ for
$n\geq m>0$. From \eqref{9a.1} we infer:
$$\| U(n,m)|_{\im P_m^+}\|=\|U(n,m)|_{\im P_m}\| 
\leq Me^{-\alpha (n-m)},\quad n\geq m>0.$$
The  matrix representation \eqref{1B.1} 
of the operator $A=U(n,m)|_{\ker \, P^+_m}$ in the decompositions
$\ker  P^+_m=\ker  P_m\oplus W_m$
and $\ker  P^+_n=\ker  P_n\oplus W_n$ is diagonal by \eqref{9a.2} and
Lemma~\ref{SL5.1}(iii) with the  invertible diagonal blocks $U(n,m)|_{\ker P_m}$ and $U(n,m)|_{W_m}$. 
 Then the operator $U(n,m)|_{\ker \, P_m}$ is invertible; its inverse
satisfies the estimate in Proposition \ref{P9.1}(iii). 
The operator $U(n,m)|_{W_m}$ satisfies \eqref{S5.2}. 
Thus, we have $\|(U(n,m)|_{\ker \, P^+_m})^{-1}\| \leq
Me^{-\alpha (n-m)}$ for $n\geq m>0$.

Next, consider $n=0$. Recall that $X'_0$ is a direct complement
of $X_0$ in $\Y0$, and
that $X_0\subset \ker \, P_0$ by \eqref{12.3} and $\ker P_0\subset\Y0$
by Proposition \ref{P9.1}.
Denote $\tilde{X}_0=X'_0\cap\ker P_0$. For each $x\in\ker \, P_0$ use the direct
sum decomposition $\Y0=X_0\oplus X'_0$ to write $x=x_0+x'_0$ with unique
$x_0\in X_0$, $x'_0\in X'_0$. Then $x'_0=x-x_0\in\ker P_0$ and thus
$x'_0\in\tilde{X}_0$. So, $\tilde{X}_0$ is a direct complement of $X_0$ in
$\ker \, P_0$, that is, $X_0\oplus \tilde{X}_0=\ker \, P_0$.  We claim that
\begin{equation}\label{S7A.1} U(1,0):\tilde{X}_0\to \ker P_1\text{ is an isomorphism.}\end{equation}
Indeed, if $x\in \ker P_1$ then, by the surjectivity of the node operator $N(1,0)$ from Proposition~\ref{P9.1}(iv) there exists $y\in 
\ker \, P_0$ so that
$N(1,0)y=(I-P_1)U(1,0)y=x$. Use the direct sum decomposition $\ker \, P_0=X_0\oplus \tilde{X}_0$ to write $y=y_0+\tilde{y}_0$, where $y_0\in X_0$, $\tilde{y}_0\in
\tilde{X}_0$. Since $\ker N(1,0)=X_0$, we have
$x=N(1,0)y=N(1,0)\tilde{y}_0=(I-P_1)U(1,0)\tilde{y}_0$. Since $\tilde{y}_0\in
\tilde{X}_0\subset\ker P_0$, we have $P_0\tilde{y}_0=0$. But $\tilde{y}_0\in
\tilde{X}_0\subset X'_0$, and \eqref{12.2} then implies
$0=U(1,0)P_0\tilde{y}_0=P_1U(1,0)\tilde{y}_0$. Thus, $U(1,0)\tilde{y}_0\in\ker
\, P_1$, and $U(1,0)\tilde{y}_0=(I-P_1)U(1,0)\tilde{y}_0=x$.
Therefore, $U(1,0):\tilde{X}_0\to \ker P_1$ is surjective. Next, if $U(1,0)\tilde{y}_0=0$ for some $\tilde{y}_0\in \tilde{X}_0\subset 
\ker P_0$, then
$N(1,0)\tilde{y}_0=0$. Since $\ker \, N(1,0)=X_0$ by 
Proposition~\ref{P9.1}(iv), we have $\tilde{y}_0\in X_0$
and thus $\tilde{y}_0=0$ since $X_0\cap \tilde{X}_0=\{0\}$.
This proves \eqref{S7A.1}.

Define a projection $P^+_0$ on $X$ such that 
\begin{equation}\label{SN7A.5}\im P^+_0=\im P_0\oplus X_0\text{ 
and }\ker P^+_0=Y_0\oplus \tilde{X}_0\end{equation}
so that $X=\im P^+_0\oplus \ker P^+_0$ by
\eqref{6.1} and $\Y0=\ker P_0\oplus \im P_0$ by Proposition~\ref{P9.1}. 
Recall that $\im P^+_1=\im P_1$ and $\ker P^+_1=W_1\oplus \ker P_1$, 
see \eqref{S7N.1}.
Note that we have $U(1,0)(X_0)\subseteq X_1\subset \im P_1$ by Lemma~\ref{L4.1}(ii) and \eqref{12.3}. 
Also, \begin{equation}\label{SN7B.1}U(1,0)(\im P_0)\subset \im
P_1.\end{equation} Indeed, using Proposition~\ref{P9.1}(ii), 
we have that if $x=P_0x$ then $U(1,0)x=U(1,0)P_0x=P_1U(1,0)y'_0\in \im P_1$. 
Thus, $U(1,0)\im P^+_0\subset\im P^+_1$. Also, we have that
$U(1,0)(Y_0) = W_1\subset \ker P^+_1$ by Lemma~\ref{SL5.1}(iii) 
and $U(1,0)(\tilde{X}_0) = \ker P_1 \subset \ker P^+_1$ by claim \eqref{S7A.1}. This proves
$U(1,0) (\ker P^+_0) \subset \ker P^+_1$ and $U(1,0)P^+_0=P^+_1U(1,0)$.

For $n\geq 2$ and $x\in \im P^+_0$ we have $\| U(n,0)x\|=\| U(n,1)U(1,0)x\| \leq Me^{-\alpha (n-1)}\| U(1,0)x\|\leq M'e^{-\alpha n}\| x\|$ 
because $U(1,0)x\in \im
P^+_1$. Also, the restriction $U(n,0)|_{\ker P^+_0}=U(n,1)|_{\ker P^+_1}U(1,0)|_{\ker P^+_0}$ is invertible from 
$\ker P^+_0$ to $\ker P^+_n$. Indeed, $U(n,1)|_{\ker
P^+_1}: \ker P^+_1\to \ker P^+_n$ is invertible by the proof of dichotomy for $n\geq 1$. 
Also, $U(1,0)|_{\ker P_0^+}:\ker P^+_0\to \ker P^+_1$ is a direct sum
of two operators, $U(1,0)|_{Y_0}:Y_0\to W_1$ and $U(1,0)|_{\tilde{X}_0}:\tilde{X}_0\to \ker P_1$. The first operator is invertible by Lemma~\ref{SL5.1}(iv)
and the second operator is invertible by claim \eqref{S7A.1}. Exponential estimates for $\| (U(n,0)|_{\ker P^+_0})^{-1}\|$ follow from the estimates for $\|
(U(n,1)|_{\ker P^+_1})^{-1}\|$. 
\hfill$\Box$

{\bf PROOF OF THEOREM \ref{FREDSA} FOR 
{\boldmath$n\le0$\unboldmath}.}
It is convenient to work on $X^*$ with the 
family $\{U(n,m)^*\}_{0\geq n\geq m}$. First, consider $n<0$.
By Proposition~\ref{PS3.1} and Lemma~\ref{SL9.1}(ii) 
we have the direct sum decomposition
$X^*=X^\bot_n\oplus W_{n,*}=\im P_{n,*}\oplus\ker P_{n,*}\oplus W_{n,*}$,
$n<0$.
Let $R_{n,*}$ be a projection on $X^*$ such that 
\begin{equation}\label{S10.N.1}\im R_{n,*}=\im P_{n,*}\text{ and }\ker R_{n,*}=
\ker P_{n,*}\oplus W_{n,*},\quad
n<0.\end{equation}
As in the proof of Theorem~\ref{FREDSA} for $n>0$, 
one checks for $0>n\geq m$ the following assertions:
\begin{align}\label{S8.1} & U(n,m)^*R_{n,*}=R_{m,*}U(n,m)^*;\\
\label{S8.2} & \| U(n,m)^*|_{\im R_{n,*}}\| \leq Me^{-\alpha (n-m)};\end{align}
the restriction $U(n,m)^*|_{\ker R_{n,*}}:\ker R_{n,*}\to 
\ker R_{m,*}$ is invertible, and
\begin{equation}\label{S8.3} \| (U(n,m)^*|_{\ker R_{n,*}})^{-1}\| 
\leq Me^{-\alpha (n-m)}.\end{equation}
Next, consider $n=0$. Let $X'_{0,*}$ be a direct complement of $X_{0,*}$
in $X_0^\bot$ and recall that $X_{0,*}\subset\ker P_{0,*}$ by \eqref{S3.6}.
Denote $\tilde{X}_{0,*}=X'_{0,*}\cap\ker P_{0,*}$, so that $\ker
P_{0,*}=X_{0,*}\oplus
\tilde{X}_{0,*}$. 
Define a projection $R_{0,*}$ on $X^*$ as follows:
\begin{equation}\label{SN13.1} \im R_{0,*}=\im P_{0,*}\oplus X_{0,*},
\quad\ker R_{0,*}
=Y_{0,*}\oplus\tilde{X}_{0,*}.\end{equation}
We now prove that assertions \eqref{S8.1}-\eqref{S8.3} hold for $0\geq n\geq m$ (cf. the corresponding part of the proof of Theorem~\ref{FREDSA} for $n\geq m\geq
0$). Recall from \eqref{S10.N.1} that
\begin{equation}\label{SN13.2} \im R_{-1,*}=\im P_{-1,*},
\quad \ker R_{-1,*}=\ker P_{-1,*}\oplus W_{-1,*}.\end{equation}
Note that $U(0,-1)^*( X_{0,*}) \subset X_{-1,*}\subset \im P_{-1,*}$ by Lemma~\ref{L4.1}(iii) and \eqref{S3.6}. 
Also, $U(0,-1)^* (\im P_{0,*}) \subset \im P_{-1,*}$ as in
\eqref{SN7B.1}. Indeed, if $\xi=P_{0,*}\xi$ then $U(0,-1)^*\xi \in 
\im P_{-1,*} $ by \eqref{SN12.3}. 
Thus, we have  $U(0,-1)^*(\im R_{0,*}) \subset \im
R_{-1,*}$. To prove $U(0,-1)^* (\ker R_{0,*}) \subset \ker R_{-1,*}$, 
we first remark (cf. \eqref{S7A.1}) that
\begin{equation}\label{SN13.3} U(0,-1)^*:\tilde{X}_{0,*}\to 
\ker P_{-1,*}\text{ is an isomorphism.}\end{equation}
The proof of \eqref{SN13.3} is identical to the proof of \eqref{S7A.1} 
and uses the
reduced node operator \eqref{SN12.5}. Lemma~\ref{SL9.1}(iii),(iv) implies that
$U(0,-1)^*:Y_{0,*}\to W_{-1,*}$ is an isomorphism. Thus, by \eqref{SN13.1}, \eqref{SN13.2}, and \eqref{SN13.3} we conclude that $U(0,-1)^*:\ker R_{0,*}\to \ker
R_{-1,*}$ is an isomorphism. So,
 $U(0,-1)^*R_{0,*}=R_{-1,*}U(0,-1)^*$. The estimates \eqref{S8.2}-\eqref{S8.3} for $0\geq n\geq m$ 
(with, generally, new $M$)
follow from the estimates for $0>n\geq m$ that have been previously
proved  in Proposition \ref{PS3.1} 
and Lemma \ref{SL5.1}.

To finish the proof of Theorem \ref{FREDSA} for $n\le0$, we denote
$P^-_n=(R_{n,*})^*$, $n\leq0$, and observe that
$\im P^-_n=\im (R_{n,*})^*=
(\ker R_{n,*})^\bot =(\im R_{n,*})^*$, and
\begin{equation}\ker P^-_n=\ker (R_{n,*})^*=(\im R_{n,*})^\bot =
(\ker R_{n,*})^*.\label{S10.N.1.2}\end{equation}
Passing to the adjoint operators in \eqref{S8.2}-\eqref{S8.3}, we have for $0\geq n\geq m$:
\begin{equation}\label{S10.N.2} \| U(n,m)|_{\im P_m^-}\| \leq Me^{-\alpha(n-m)},
\|(U(n,m)|_{\ker P^-_m})^{-1}\| \leq Me^{-\alpha(n-m)},\end{equation}
and Theorem \ref{FREDSA} for $n\le0$ is proved.\hfill$\Box$

The next statement shows that the dimension of the 
kernel and cokernel of $D$ is, in fact, equal to the dimension
of the arbitrary fiber, cf. Lemma \ref{L4.1}(i).
\begin{cor}\label{CS11.1} If $D$ is Fredholm, then
for each $n\in \mathbb{Z}$ we have
 $\dim X_n=\dim \ker D$ 
 and $\dim X_{n,*}=\dim \ker D^*$.\end{cor}

\begin{pf} Fix $x\in X_n$, and let $x_k=U(k,n)x$ for $k\geq n$. By Lemma~\ref{L4.1}(ii), $x_k\in X_k$. Using \eqref{S7N.1} and Lemma~\ref{XnNR}, for $k>\max \{n,0\}$
we have $x_k\in \im P^+_k$. Thus, $\|x_k\|\leq ce^{-\alpha k}$
for $k\ge0$. If $k<n$ then by Lemma~\ref{L4.1}(ii) there exists a 
unique $x_k\in X_k$ such that $x=U(n,k)x_k$.
Using \eqref{S10.N.1} and \eqref{S10.N.1.2}, for $k<\min \{0,n\}$ we have $\ker P^-_k=(\im R_{k,*})^\bot =
(\im P_{k,*})^\bot \supset X_k$ since $\im P_{k,*}\subset
X^\bot_k$ in Proposition~\ref{PS3.1}. 
Thus, $\| x\|=\|U(n,k)x_k\|\geq ce^{-\alpha k}\| x_k\|$ 
or $\| x_k\|\leq ce^{\alpha k}$ for $k<0$. 
Therefore, starting with an $x\in
X_n$, we  obtain an exponentially decaying as $|k|\to\infty$ sequence 
$(x_k)_{k\in \mathbb{Z}}$ such that $x_k=U(k,m)x_m$ for all $k\geq m$ in $\mathbb{Z}$. Thus, $(x_k)_{k\in
\mathbb{Z}}\in \ker D$, and we can consider a 
well-defined and injective linear map $j_n:X_n\to \ker D:
x\mapsto (x_k)_{k\in\mathbb{Z}}$.  
It is surjective by
the definition of $X_n$. Thus, $X_n$ and $\ker D$ are isomorphic. 
Similarly, $X_{n,*}$ is isomorphic to $\ker D^*$.
\hfill$\Box$\end{pf}

\section{Proof of Theorems \ref{main} and \ref{mainRef}}\label{PFMT}
In this section, in Proposition \ref{CSN14.1} we show that if $D$
is Fredholm then the discrete node operator $N(n,m)$, $n\ge m$ in
$\mathbb{Z}$, is Fredholm, and that $\ind D=\ind N(n,m)$.
Thus, Theorem \ref{FREDSA} and Proposition \ref{CSN14.1} 
in combination with Theorem \ref{FrGandD} 
yield the implication \eqref{GFr} $\Rightarrow$ 
(i) and (ii) in Theorem \ref{main}. Finally, to complete the proofs
 of Theorem \ref{main} and \ref{mainRef}, we show that 
(i) and (ii) in Theorem \ref{main} imply \eqref{GFr}.

Consider two families of projections, $\{P_n^-\}_{n\le0}$
and $\{P_n^+\}_{n\ge0}$.
For $n\geq 0\geq m$ we define the  discrete node operator $N(n,m)$ as follows:
$$N(n,m):=(I-P^+_n)U(n,m)|_{\ker P_m^-}:\ker P^-_m\to \ker P^+_n.$$
Note that $N(0,0)=(I-P^+_0)| _{\ker P^-_0}$ 
acts from $\ker P^-_0$ to $\ker P^+_0$,  and 
 $N(0,0)=(I-P^+_0)(I-P_0^-)| _{\ker P^-_0}$. 
 First, we reformulate the fact that $N(0,0)$ is Fredholm in terms of the associated Fredholm pair of subspaces.
\begin{lem} \label{REQ} If $(P^+_0,P^-_0)$ is a pair of projections on $X$, then the node operator 
$N(0,0)=(I-P^+_0)|_{\ker
P^-_0}$ is a Fredholm operator from $\ker P^-_0$ to $\ker P^+_0$ if and only if the pair of subspaces $\ker P^-_0$ and $\im P^+_0$ is
Fredholm in $X$. Moreover, $\dim \ker N(0,0)=
\alpha (\ker P^-_0,\im P^+_0)$, $\codim \im N(0,0)=\beta(\ker P^-_0,\im P^+_0)$, and $\ind
N(0,0)=\ind (\ker P^-_0,\im P^+_0)$.\end{lem}
\begin{pf} By the definition of $N(0,0)$ we have $\ker N(0,0)=\ker P^-_0\cap \im P^+_0$. 
We claim that $\im N(0,0)\oplus \im
P^+_0=\ker P^-_0+\im P^+_0$. Indeed, if $x\in\ker P_0^-$ then
$y=N(0,0)x=x-P_0^+ x\in\ker P_0^-+\im P_0^+$, and
the inclusion ``$\subset$" holds. To prove the inclusion 
``$\supset$", take a $z=x+y$ with
$x\in \ker P^-_0$ and $y\in \im P^+_0$. Then $(I-P^+_0)z=(I-P^+_0)x\in \im N(0,0)$, and $z=(I-P^+_0)z+P^+_0z\in \im N(0,0)\oplus \im
P^+_0$. Using the claim, $\im N(0,0)$ is a closed subspace in $\ker P^+_0$ if and only if $\ker P^-_0+\im P^+_0$ is a closed subspace in
$X$, and $\dim (\ker P^+_0/\im N(0,0))=\dim (X/(\im N(0,0)\oplus 
\im P^+_0)=\beta (\ker P^-_0,\im P^+_0)$ for the quotient spaces.
\hfill$\Box$\end{pf}

\begin{prop}\label{CSN14.1} If $D$ is Fredholm on
$\ell_p(\mathbb{Z};X)$, $p \in [1, \infty)$,
or on $c_0(\mathbb{Z};X)$, then the discrete node operator $N(n,m)$, $n\geq 0\geq m$,
is Fredholm. Moreover, $\dim\ker D=\dim\ker N(n,m)$,
$\codim\im D=\codim\im N(n,m)$, and $\ind D=\ind N(n,m)$.\end{prop}

\begin{pf} Consider the dichotomies
$\{ P^+_n\}_{n\geq 0}$ and $\{ P^-_n\}_{n\leq 0}$ 
for $\{ U(n,m)\}_{n\geq m}$, obtained in Theorem
\ref{FREDSA}. Note that $N(n,m)=N(n,0)N(0,0)N(0,m)$, $n\geq
0\geq m$, and that operators $N(n,0)$, $n>0$, and $N(0,m)$, $0>m$, 
are invertible. 
Thus, it suffices to prove that $N(0,0)$ is Fredholm and $\ind N(0,0)=\ind
D(=\dim X_0-\dim X_{0,*})$, see Corollary~\ref{CS11.1}. 
We know that $\im D $ is closed, and want 
to derive that $\im N(0,0)$ is closed. First, we claim that if 
$y=(I-P^+_0)x$, $x\in \ker P^-_{0}$, then $y\otimes {\bf
e}_0\in \im D$. Indeed, define $x_n=(U(0,n)|_{\ker P^-_n})^{-1}x$ for $n< 0$ and $x_n=U(n,0)P^+_0x$ for $n\geq 0$. Then for $n<0$ we have
$x_n-U(n,n-1)x_{n-1}=(U(0,n)|_{\ker P^-_0})^{-1}x-U(n,n-1)
(U(0,n-1)|_{\ker P^-_{n-1}})^{-1}x=0$. Similarly, for $n>0$ we have
$x_n-U(n,n-1)x_{n-1}=U(n,0)P^+_0x-U(n,0)P^+_0x=0$. For $n=0$ we have 
\begin{equation*}\begin{split}x_0-U(0,-1)x_{-1}&=
P^+_0x-U(0,-1)(U(0,-1)|_{\ker P^-_{-1}})^{-1}x\\
&=P^+_0 x-(I-P^-_0)x=P^+_0x-x=-y,\end{split}\end{equation*}
where we have used that $x\in \ker P^-_0$. Thus, $y\otimes {\bf e}_0\in \im D$ as claimed. Second, we claim that if $y\otimes {\bf e}_0\in \im D$ and $y\in \ker
P^+_0$, then $y\in \im N(0,0)$. Indeed, for some ${\bf x}\in \ell_p(\mathbb{Z};X)$ we have $D{\bf x}=y\otimes {\bf e}_0$. 
Thus $0=x_n-U(n,0)x_0$ for $n > 0$.
This implies $x_0\in \im P^+_0$. Also, $0=x_{-1}-U(-1,n)x_n$ for $n\leq -1$.  Therefore $x_{-1}\in \ker P^{-}_{-1}$ and $U(0,-1)x_{-1}\in \ker P^-_0$. Finally,
$y=x_0-U(0,-1)x_{-1}$ yields that $y=(I-P^+_0)y=(I-P^+_0)x_0-(I-P^+_0)
U(0,-1)x_{-1}=-(I-P^+_0)U(0,-1)x_{-1}\in \im N(0,0)$ since 
$x_0\in \im P^+_0$ and
$U(0,-1)x_{-1}\in \ker P^-_0$, and the second claim is proved. Now assume $y=\lim_{j\to \infty}y^{(j)}$, 
where $y^{(j)}\in \im N(0,0)$. By the first claim
$y^{(j)}\otimes {\bf e}_0\in \im D, j \in \mathbb N$. Since $\im D$ is closed, $y\otimes {\bf e}_0=\lim_{j\to \infty}y^{(j)}\otimes {\bf e}_0 \in \im D$. Since $\im N(0,0)\subset
\ker P^+_0$, we also have $y\in \ker P^+_0$. By the second claim $y\in \im N(0,0)$, and thus $\im N(0,0)$ is closed.

Next, we prove the formulas for the defect numbers.
We have $\ker N(0,0)=\ker P^-_0\cap \im P^+_0$. 
Thus, if $x\in \ker N(0,0)$ then $\| x_n\|\leq ce^{-\alpha
n}$, for $x_n=U(n,0)x$, $n\geq 0$, since $x\in \im P^+_0$. 
Also, $\| x_n\|\leq ce^{\alpha n}$, $n<0$, for the sequence 
$(x_n)_{n<0}$ such that $x=U(0,n)x_n$, $n<0$, since $x\in\ker P_0^-$.
Thus, with this choice of $x_n$ we have $x_n=U(n,m)x_m$ for all
$n\geq m$, and $(x_n)_{n\in \mathbb{Z}}\in \ker D$. Thus, $x\in X_0$. 
On the other hand,
$$\ker P^-_0=(\im R_{0,*})^\bot =(\im P_{0,*}\oplus X_{0,*})^\bot =(\im P_{0,*})^\bot \cap (X_{0,*})^\bot$$
by \eqref{S10.N.1.2} and \eqref{SN13.1}. 
Since $X_0\subset \im P^+_0\subset
X_{0,*}^\bot$ by \eqref{SN7A.5} and $\im P_{0,*}\subset X^\bot_0$
by Proposition \ref{PS3.1}, we have
$\ker N(0,0)=\im P^+_0\cap [X^\bot_{0,*}\cap (\im P_{0,*})^\bot ]=
\im P^+_0\cap (\im P_{0,*})^\bot\supset X_0$.
So, $\ker N(0,0)=X_0$, and $\dim \ker N(0,0)=\dim X_0$. Further,
$N(0,0)^*=(I-P^-_0)^*(I-P^+_0)^*$  is  an
operator acting from $(\ker P^+_0)^*=\ker (P^{+}_0)^*$ to 
$\ker (P^{-}_0)^*=(\ker P^-_0)^*$, and $\ker N(0,0)^* = \im (P^{-}_0)^* \cap \ker (P^+_0)^*.$ 
A similar argument yields  $\dim N(0,0)^*=X_{0,*}$.\hfill$\Box$\end{pf}

\noindent{\bf PROOFS OF THEOREMS \ref{main} AND \ref{mainRef}.}
 Assume ${\bf G}$ is Fredholm. Then $D$ is Fredholm by Theorem~\ref{FrGandD}. By
Theorem~\ref{FREDSA} there exist discrete dichotomies on $\mathbb{Z}_+$ and $\mathbb{Z}_-$. By Lemma~\ref{DCDich}, 
there exist dichotomies $\{ P^+_t\}_{t\geq 0}$
and $\{ P^-_t)_{t\le 0}$. This proves (i') in Theorem~\ref{mainRef} and,
therefore, (i) in Theorem~\ref{main} for $a=0=b$. 
By Proposition \ref{CSN14.1} we also have
that $N(0,0)$ is Fredholm, and, using formulas for the defect numbers 
and index from Theorem~\ref{FrGandD}, 
we derive (ii') in Theorem~\ref{mainRef} and, by Lemma \ref{REQ},
(ii) in Theorem~\ref{main} for $a=0=b$, 
and the required formulas for the defect numbers and the index.
It remains to prove that (i) and (ii) in Theorem~\ref{main} imply \eqref{GFr},
see \cite[Thm.4]{BaskDAN02}, and also \cite[Thm.8]{Ba} 
for the proof in the case when $a=-1$ and $b=0$. 
We will present a proof, 
different form \cite{Ba}, as well as from the corresponding proofs 
in \cite{AM1,BAG,Lin,Pa,SS2,SS1}
given in particular cases. Our proof is 
based on the following abstract fact from \cite[p. 23]{Spitk}.

\begin{lem}~\label{LSp} Assume that a bounded linear operator $A$ acting 
on a direct sum $\mathcal{X}_1\oplus \mathcal{X}_2$ of two Banach spaces has the
following triangular representation:
\begin{equation}\label{NN7.1} A=\begin{bmatrix} A_{11} & 0\\ A_{21} & 
A_{22} \end{bmatrix},\text{ where } A_{11} 
\in \mathcal{L}(\mathcal{X}_1),\, A_{21}\in
\mathcal{L}(\mathcal{X}_1,\mathcal{X}_2),\,
 A_{22}\in \mathcal{L}(\mathcal{X}_2).\end{equation}
Then $A$ is Fredholm if and only if the following assertions hold.
\begin{enumerate}\item[(i)] $\im A_{11}$ is closed , and $\codim \im A_{11}<\infty$;
\item[(ii)] $\im A_{22}$ is closed, and $\dim \ker A_{22}<\infty$;
\item[(iii)]  
If  $\mathcal{L}_1:=\{x\in \mathcal{X}_1:x\in \ker A_{11}$ 
and $A_{21}x\in \im A_{22}\}$ 
then $\dim\mathcal{L}_1$ is finite;
\item[(iv)]  
If $\mathcal{L}_2:=\im A_{22}+A_{21} (\ker A_{11})$ 
then  $\codim\mathcal{L}_2$ in $\mathcal{X}_2$ is finite. \end{enumerate}
If (i)-(iv) holds, then $\dim \ker A=\dim \ker A_{22}
+\dim
\mathcal{L}_1$ and $\codim \im A=\codim \im A_{11}+\codim \mathcal{L}_2$.\end{lem}

By Theorem~\ref{FrGandD}, it suffices to prove that $D$ is Fredholm
provided (i) and (ii) in Theorem \ref{main} hold. We will present the proof
for the $\ell_p$--case, the $c_0$--case is similar. 
Passing to $[a]-1$ and $[b]+1$, if needed, 
where $[\cdot]$ is the integer part, we may
assume that:
(1)\,\, $a,b\in\mathbb{Z}$ in Theorem~\ref{main}; 
(2)\,\, the discrete evolution family $\{ U(n,m)\}_{n\geq m}$, 
$n,m\in \mathbb{Z}$, has dichotomies $\{ P^-_n\}_{n\leq a}$ and $\{
P^+_n\}_{n\geq b}$; and (3)\,\, the discrete node operator 
$N(b,a)=(I-P_b^+)U(b,a)|_{\ker P^-_a}$ is a Fredholm operator from 
$\ker P^-_a$ to $\ker P^+_b$. 
First,
for $A=D$ consider representation
\eqref{NN7.1} for $\ell_p(\mathbb{Z};X)=\mathcal{X}_1\oplus \mathcal{X}_2$ with 
$\mathcal{X}_1=\ell_p(\mathbb{Z}\cap (-\infty ,b];X)$ and
$\mathcal{X}_2=\ell_p(\mathbb{Z}\cap [b+1,\infty);X)$. Then $A_{11}=D^{-}_b$, 
where $D^-_b=D|_{\ell_p(\mathbb{Z}\cap (-\infty, b];X)}$, $A_{22}=D^+_b$, 
where
$D^+_b:(x_n)_{n\geq b+1}\mapsto (x_{b+1},x_{b+2}-U(b+2,b+1)x_{b+1},\ldots)$, 
and $A_{21}=D^\pm_b$, where
$D^\pm_b:(x_n)_{n\leq b}\mapsto (-U(b+1,b)x_b,0,\ldots )$. Therefore,
\begin{align} 
\mathcal{L}_1&=
\{ (x_n)_{n\leq b}:(x_n)_{n\leq b}\in \ker D^-_b\text{ and } \nonumber\\
&\quad \hskip3cm (-U(b+1,b)x_b,0,\ldots )\in \im D^+_b\},\label{NN8.1}\\
\begin{split}\mathcal{L}_2&=\{ (x_n)_{n\geq b+1}+(-U(b+1,b)x_b,0,\ldots ):\\
&\quad \hskip3cm (x_n)_{n\geq b+1}\in \im D^+_b\text{ and } (x_n)_{n\leq b}\in \ker
D^-_b\}.\label{NN8.2}\end{split}\end{align}
We will need a version of \cite[Cor.1]{Ba}.
For a sequence $(x_n)_{n\ge b+2}$ denote
\begin{equation}\label{NN8.4}
x'_{b+1}=-\sum^\infty_{k=1}(U(b+1+k,b+1)|_{\ker P^+_{b+1}})^{-1}
(I-P^+_{b+1+k})x_{b+1+k}.
\end{equation}
The series in \eqref{NN8.4} converges by the unstable dichotomy estimate.

\begin{lem}\label{LNN8.1} The operator $D^+_b$ is left-invertible on $\ell_p(\mathbb{Z}\cap [b+1,\infty);X)$, and
\begin{equation}\label{NN8.3} \im D^+_b=
\{ (x_n)_{n\geq b+1}:(I-P^+_{b+1})x_{b+1}=x'_{b+1}\}.\end{equation}
\end{lem}

\begin{pf} To construct $(D^+_b)^{-1}$, the left inverse for $D^+_b$, note that $D^+_b=I-T^+_b$, 
where $T^+_b:(x_n)_{n\geq b+1}\mapsto
(0,U(b+2,b+1)x_{b+1},\ldots )$. Decompose $T^+_b=T^+_{b,s}\oplus T^+_{b,u}$, where $T^+_{b,s}$, 
respectively, $T^+_{b,u}$, is the restriction of $T^+_b$ on the
subspace of sequences
$(x_n)_{n\geq b+1}$ {\bf }from $\ell_p(\mathbb{Z}\cap [b+1,\infty);X)$ such that $x_n\in \im P^+_n$, respectively, $x_n\in \ker P^+_n$, $n\geq b+1$.
Then
$T^+_{b,u}$ is left invertible with the left inverse $(T^+_{b,u})^{-1}:(x_n)_{n\geq b+1}\mapsto ( U(n+1,n)|_{\ker P^+_n})^{-1}x_{n+1})_{n\geq b+1}$. By the
dichotomy assumption, $\sprad (T^+_{b,s})<1$ and $\sprad ((T^+_{b,u})^{-1})<1$, and thus
$(D^+_b)^{-1}=\sum^\infty_{k=0}(T^+_{b,s})^k-
\sum^\infty_{k=1}(T^+_{b,u})^{-k}$. 
A calculation shows that $D^+_b(D^+_b)^{-1}$ maps 
a sequence $(x_n)_{n\geq b+1}$
to the sequence
$(P^+_{b+1}x_{b+1}+x'_{b+1},
 x_{b+2},\ldots )$, see \eqref{NN8.4}.
Since $\im D^+_b=\im (D^+_b(D^+_b)^{-1})$, we obtain \eqref{NN8.3}.\hfill$\Box$

Using the decomposition $\ell_p(\mathbb{Z}\cap (-\infty ,b];X)=\mathcal{X}_1\oplus \mathcal{X}_2$, where $\mathcal{X}_1=\ell_p(\mathbb{Z}\cap (-\infty ,a-1];X)$
and $\mathcal{X}_2=\ell_p(\mathbb{Z}\cap [a,b];X)$, consider  representation \eqref{NN7.1} for $A=D^-_b$. We now have 
$A_{11}=D^-_{a-1}=D|_{\ell_p(\mathbb{Z}\cap
(-\infty ,a-1];X)}$, and also $A_{22}=D_{a,b}$, 
where $D_{a,b}:(x_n)_{a\leq n\leq b}\mapsto (x_a,x_{a+1}-U(a+1,a)x_a,\ldots ,x_b-U(b,b-1)x_{b-1})$. In the
representation $\ell_p(\mathbb{Z}\cap [a,b];X)=X\oplus \ldots \oplus X$ 
($(b-a)-$times) the operator $D_{a,b}$ is lower-triangular  with identities on the diagonal and, hence, invertible. Using
dichotomy
$\{P^-_n\}_{n\leq a-1}$, similarly to the proof of Lemma~\ref{LNN8.1}, we conclude that $D^-_{a-1}$ is right-invertible. Since $D^-_b$ is lower triangular with the
diagonal blocks
$D^-_{a-1}$ and $D_{a,b}$, it follows that $D^-_b$ is right-invertible. 
This and Lemma~\ref{LNN8.1} imply that for the triangular representation 
\eqref{NN7.1} of
$D$ both assertions (i) and (ii) hold. Thus, to conclude that $D$ is Fredholm,
 it remains to prove that $\dim \mathcal{L}_1<\infty$ and $\codim
\mathcal{L}_2<\infty$ for $\mathcal{L}_1$ and $\mathcal{L}_2$ in \eqref{NN8.1}-\eqref{NN8.2}. As soon as this is proved, $\dim \ker D=\dim \mathcal{L}_1$ and
$\codim \im D=\codim \mathcal{L}_2$.

To handle $\mathcal{L}_1$, remark that $(-U(b+1,b)x_b,0,\ldots )\in \im D^+_b$ if and only if there exists 
a $(y_n)_{n\geq b+1}\in \ell_p(\mathbb{Z}\cap
[b+1,\infty ); X)$ such that $y_n=-U(n,b)x_b$, $n\geq b+1$. Using the dichotomy $\{ P^+_n\}_{n\geq b}$, this is equivalent to $x_b\in \im P^+_b$. On the other hand,
$(x_n)_{n\leq b}\in \ker D^-_b$ means that $x_n=U(n,m)x_m$ for all $m\leq n\leq b$. In particular, $x_b=U(b,a)x_a$, and $x_a=U(a,n)x_n$ for all $n\leq a$. 
Using
the dichotomy $\{ P^-_n\}_{n\leq a}$, we infer $x_a\in \ker P^-_a$. Thus,
$$ \dim \mathcal{L}_1=\dim \{ x\in \ker P^-_a:U(b,a)x \in \im P^+_b\}=
\dim \ker N(b,a)<\infty.$$
To handle $\mathcal{L}_2$, let $Z$ denote any direct complement of $\im N(b,a)$, such that 
$\ker P^+_b=\im N(b,a)\oplus Z$, and let $[(x_n)_{n\geq
b+1}]_{\mathcal{L}_2}$ for any $(x_n)_{n\geq b+1}\in \ell_p(\mathbb{Z}\cap [b+1,\infty );X)$ denote the equivalence class in the quotient space
$\ell_p(\mathbb{Z}\cap [b+1,\infty);X)/\mathcal{L}_2$. 
By Lemma~\ref{LNN8.1} we have $(P^+_nx_n)_{n\geq b+1}\in \im D^+_b\subset \mathcal{L}_2$. Thus,
$[(x_n)_{n\geq b+1}]_{\mathcal{L}_2}=
[((I-P_n^+)x_n)_{n\geq b+1}]_{\mathcal{L}_2}$. 
Using \eqref{NN8.1}, by Lemma~\ref{LNN8.1} we infer  
$(x'_{b+1},(I-P^+_{b+2})x_{b+2},\ldots )\in\im D^+_b\subset \mathcal{L}_2$,
 so, $[(x_n)_{n\geq b+1}]_{\mathcal{L}_2}=
[(y_{b+1},0, \ldots )]_{\mathcal{L}_2}$, where we
denote $y_{b+1}=(I-P^+_{b+1})x_{b+1}-x'_{b+1}$. Note that $y_{b+1}\in
\ker P^+_{b+1}$, and find the unique $y_b\in \ker P^+_b$ such that $y_{b+1}=U(b+1,b)|_{\ker P^+_b}y_b$. Using the decomposition $\ker P^+_b=\im N(b,a)\oplus Z$,
find the unique representation $y_b=y+z$, where $y\in \im N(b,a)$ and $z\in Z$. Since $y\in \im N (b,a)$, there is an $x_a\in \ker P^-_a$ such that $y=U(b,a)x_a$.
Using the dichotomy $\{ P^-_n\}_{n\leq a}$, set $x_n=(U(a,n)|_{\ker P^-_n})^{-1}x_a$ for $n\leq a$. Also, define $x_n=U(n,a)x_a$ for $n\in [a,b]$. Then
$(x_n)_{n\leq b}\in \ell_p(\mathbb{Z}\cap (-\infty ,b];X)$ and $x_n=U(n,m)x_m$ for all $m\leq n\leq b$. Thus, $(x_n)_{n\leq b}\in \ker D^-_b$. Also, $y=x_b$. By
\eqref{NN8.2} then $[(-U(b+1,b)y,0,\ldots )]_{\mathcal{L}_2}=[(U(b+1,b)z,0,\ldots )]_{\mathcal{L}_2}$. 
As a result, we have a well-defined map 
$j:{\bf x}=[(x_n)_{n\geq b+1}]_{\mathcal{L}_2}\mapsto z$ 
from $\ell_p(\mathbb{Z}\cap [b+1,\infty))/ \mathcal{L}_2$ 
to $Z\cong\ker P^+_b/ \im N(b,a)$ such
that $ [(x_n)_{n\geq b+1}]_{\mathcal{L}_2}=[(U(b+1,b)z,0,
\ldots )]_{\mathcal{L}_2}$ with $j{\bf x}=z.$ 
It follows that $j$ is injective. It is surjective, since if
$z\in Z$ then ${\bf x}=[(U(b+1,b)z,0,\ldots )]_{\mathcal{L}_2}$ 
satisfies $j{\bf x}=z$. \hfill$\Box$\end{pf}

\section{Differential and Difference Operators}\label{SFDD}
In this section we prove Theorem~\ref{FrGandD} and Lemma \ref{DCDich}.
The proof is given for the case of $L_p(\mathbb{R};X)$, $p\in[1,\infty)$, 
the case of $C_0(\mathbb{R};X)$ is similar. 
Fix a continuous 1-periodic function $\alpha:\mathbb{R}\to \mathbb{R}$ such that $\alpha (0)=\alpha (1)=0$ and
$\int^1_0\alpha (s)ds=0$, and recall notation 
${\bf x}=(x_n)_{n\in\mathbb{Z}}$.
Define bounded linear operators $R:L_p(\mathbb{R};X)\to \ell_p(\mathbb{Z};X)$ and 
$S:\ell_p(\mathbb{Z}; X)\to L_p(\mathbb{R};X)$ as follows:
$$Rf=\left(-\int^n_{n-1}U(n,s)f(s)ds\right)_{n\in \mathbb{Z}},\,
 (S{\bf x})(t)=\alpha (t)U(t,n)x_n,\, t\in [n,n+1].$$

\begin{lem}\label{LNN5.1} \begin{enumerate}\item[(i)] 
If ${\bf y}=D{\bf x}$ then ${\bf G}u=S{\bf y}$ for some $u\in \dom {\bf G}$;
\item[(ii)] if $S{\bf y}={\bf G}u$ for some $u\in \dom {\bf G}$ then ${\bf y}=D{\bf x}$ for some 
${\bf x}\in \ell_p$;
\item[(iii)] if $f={\bf G}u$ for some $u\in \dom {\bf G}$, then $Rf=D{\bf x}$ with ${\bf x}=(u(n))_{n\in \mathbb{Z}}$;
\item[(iv)] if $Rf=D{\bf x}$ for some ${\bf x}\in\ell_p$, then $f={\bf G}u$ for some $u\in \dom {\bf
G}$.\end{enumerate}\end{lem}

\begin{pf} { (i)} Define $u(t)=U(t,n)(y_n-x_n)-\int ^t_nU(t,s)S{\bf y}(s) \, ds$ for $t\in [n,n+1]$. A direct but tedious calculation 
similar to \cite[p.117]{ChLa99} shows
that $u\in L_p(\mathbb{R};X)\cap C_0(\mathbb{R};X)$ 
and satisfies \eqref{DGG} with $f=S{\bf y}$. Thus ${\bf G}u=S{\bf y}$.

{ (ii)} For $u\in L_p(\mathbb{R};X)\cap C_0(\mathbb{R};X)$ satisfying \eqref{DGG} with $f=S{\bf y}$ we have for $t=n+1$ and $\tau =n$:
\begin{equation*}\begin{split} u(n+1)&=U(n+1,n)u(n)-\int^{n+1}_nU(n+1,s)\alpha (s) U(s,n)y_n \, ds\\
&=U(n+1,n)u(n)-U(n+1,n)y_n,\quad n\in \mathbb{Z}.\end{split}\end{equation*}
Thus, ${\bf y}=D(y_n-u(n))_{n\in \mathbb{Z}}$.

{ (iii)} Since $u$ and $f$ satisfy \eqref{DGG}, letting $t=n$ and
$\tau =n-1$, we have that
$-\int^n_{n-1}U(n,s)f(s) \, ds=u(n)-U(n,n-1)u(n-1),\quad n\in \mathbb{Z}.$

{ (iv)} For ${\bf x}=(x_n)_{n\in \mathbb{Z}}$ such that $Rf=D{\bf x}$ define 
$$u(t)=U(t,n)x_n-\int^t_nU(t,s)f(s)ds,\quad t\in [n,n+1],\quad n\in \mathbb{Z}.$$
A calculation similar to \cite[p. 117]{ChLa99} again shows that $u\in L_p(\mathbb{R};X)\cap C_0(\mathbb{R};X)$, and that $u$ and $f$ satisfy \eqref{DGG}.
Thus, ${\bf G}u=f$.\end{pf}

We now claim that $\im {\bf G}$ is closed if and only if $\im D$ is closed. Assume that $\im D$ is closed, and consider any sequence $f^{(k)}={\bf G}u^{(k)}\in \im
{\bf G}$ such that $\lim_{k\to \infty}f^{(k)}=f$ in $L_p(\mathbb{R};X)$. Using Lemma~\ref{LNN5.1}(iii) we have $Rf^{(k)}=D(u^{(k)}(n))_{n\in \mathbb{Z}}\to Rf$,
$k\to \infty$. Since $\im D$ is closed, $Rf\in \im D$ and thus $f\in \im {\bf
G}$ by Lemma~\ref{LNN5.1}(iv). Conversely, assume that $\im {\bf G}$ is closed, and consider
any sequence ${\bf y}^{(k)}=D{\bf x}^{(k)}\in \im D$ such that $\lim_{k\to \infty}{\bf y}^{(k)}={\bf y}$ in 
$\ell_p$. Using Lemma~\ref{LNN5.1}(i),
we have $S{\bf y}^{(k)}={\bf G}u^{(k)}\to S{\bf y}$ for some $u^{(k)}\in \dom {\bf G}$. Since $\im {\bf G}$ is closed, $S{\bf y}\in \im {\bf G}$ and thus ${\bf
y}\in \im D$ by Lemma~\ref{LNN5.1}(ii). This proves the claim.

Define a linear map, $B$,
by $(B{\bf x})(t)=U(t,n)x_n$, $t\in [n,n+1)$, $n\in \mathbb{Z}$, where ${\bf x}=(x_n)_{n\in
\mathbb{Z}}$. According to \eqref{DGG}, $u\in \ker {\bf G}$ if and only if $u\in L_p(\mathbb{R};X)\cap C_0(\mathbb{R};X)$ and $u(t)=U(t,\tau )u(\tau )$ for all
$t\geq \tau $ in $\mathbb{R}$. By \eqref{2.7}, $B$ is an injective map from $\ker D$ to $\ker {\bf G}$. If $u\in \ker {\bf G}$ then $B(u(n))_{n\in \mathbb{Z}}=u$
shows that $B$ is surjective. Thus, $\ker D$ and $\ker {\bf G}$ are isomorphic, and $\dim \ker {\bf G}=\dim \ker D$.

Finally, we show that if $\im {\bf G}$ (equivalently, $\im D$) 
is closed, then 
$\dim \hat{L}_p=\dim \hat{\ell}_p$ 
for the quotient spaces $\hat{L}_p:=\{
[f]=\{f+g:g\in \im {\bf G}\}:f\in L_p\}$ and 
$\hat{\ell}_p:=\{ [{\bf y}]=\{ {\bf y}+{\bf z}:{\bf z}\in \im D\}:
{\bf y}\in \ell_p\}$.
Indeed, define the operator $\hat{R}:\hat{L}_p\to \hat{\ell}_p$,  by the rule $\hat{R}[f]=[Rf]$. Since $g\in \im {\bf G}$ implies $Rg\in \im D$ by
Lemma~\ref{LNN5.1}(iii), if $h=f+g\in [f]$, $g\in \im {\bf G}$, then $Rh=Rf+Rg\in [Rf]$, and $\hat{R}$ is well-defined. If $\hat{R}[f]=0$, then $Rf\in \im D$ and,
by Lemma~\ref{LNN5.1}(iv) we have $f\in \im {\bf G}$ and thus $[f]=0$. So, $\hat{R}$ is injective.
 Fix ${\bf y}=(y_n)_{n\in \mathbb{Z}}\in \ell_p$,
and let $f=-S{\bf y}$. Then
$$(Rf)_n=\int^n_{n-1}U(n,s)\alpha (s)U(s,n-1)y_{n-1} \, ds=y_n-(D{\bf y})_n.$$
So, ${\bf y}=Rf+D{\bf y}$. Then $[{\bf y}]=[Rf]=\hat{R}[f]$, and $\hat{R}$ is surjective. 
Thus, $\hat{L}_p$ and $\hat{\ell}_p$ are isomorphic.
\hfill$\Box$

\noindent{\bf PROOF OF LEMMA~\ref{DCDich}.} We give the proof of 
the "only if"  part for $\mathbb{R}_+$, arguments for $\mathbb{R}_-$ 
are similar. Due to the dichotomy estimates for the family $\{
U(n,m)\}_{n\geq m\geq 0}$, we claim that it suffices to construct $\{P^+_t\}_{t\geq 0}$ such that $U(t,\tau)P^+_\tau=P^+_tU(t,\tau)$ and $U(t,\tau)|_{\ker
P_\tau^+}:\ker P^+_\tau \to \ker P^+_t$ is an isomorphism for all $t\geq \tau \geq 0$. Indeed, assume that the claim is proved. Then the stable exponential
dichotomy estimate for $\{ U(t,\tau )\}_{t\geq \tau \geq 0}$ follows directly from the stable dichotomy estimate for $\{ U(n,m)\}_{n\geq m\geq 0}$ since
$\sup_{0\leq t-\tau
\leq 1}\| U(t,\tau )\|<\infty$. To obtain the unstable dichotomy estimate for
$\{U(t,\tau )\}_{t\geq \tau \geq 0}$, note that if $n+1\geq t \geq n\geq m\geq \tau \geq m-1\geq 0$ then
\begin{equation}\label{InVer} (U(t,\tau )|_{\ker P^+_\tau} )^{-1}=(U(m,\tau )|_{\ker P^+_\tau})^{-1}(U(n,m)|_{\ker P^+_m})^{-1}(U(t,n)|_{\ker
P^+_n})^{-1}.\end{equation}
But $(U(t,n)|_{\ker P^+_n})^{-1}=(U(n+1,n)|_{\ker P^+_n})^{-1}(U(n+1,t)|_{\ker P_t})$. Using the unstable dichotomy estimate for $\{U(n,m)\}_{n\geq m\geq
0}$, and the fact that $\sup\{\| U(n+1,t)\|:n\in \mathbb{Z}_+,t\in [n,n+1]\}<\infty$, we have that $\sup\{\|U(t,n)|_{\ker P^+_n})^{-1}\| :n\in \mathbb{Z}_+,t\in
[n,n+1]\}<\infty$ and, similarly, that
$\sup\{
\|(U(m,\tau )|_{\ker P^+_\tau})^{-1}\| :m\in \mathbb{Z}_+,m\geq 1,\tau \in
[m-1,m]\}<\infty$. Now \eqref{InVer} 
implies the unstable dichotomy estimate for $\{
U(t,\tau )\}_{t\geq
\tau \geq 0}$. To prove the claim, fix $t_0\in \mathbb{R}$ so that $t_0\in [n,n+1)$ for some $n\in \mathbb{Z}_+$, 
and define subspaces $X_s(t_0)=\{ x\in
X:U(n+1,t_0)x\in
\im P^+_{n+1}\}$ and $X_u(t_0)=U(t_0,n)( \ker P^+_n)$. Using the unstable dichotomy estimate for $\{ U(n,m)\}_{n\geq m\geq 0}$, for each $x\in \ker P^+_n$ we have $\|
U(n+1,t_0)\| \| U(t_0,n)x\| \geq \| U(n+1,n)x\| \geq M^{-1}e^\alpha \| x\|$. Thus, $U(t_0,n):\ker P^+_n\to X_u(t_0)$ is an isomorphism,  and $X_u(t_0)$ is closed.
Also, $U(t_1,t_0):X_u(t_0)\to X_u(t_1)$ is an isomorphism for all $t_1\geq t_0$ 
in $\mathbb{R}_+$. If $x\in X_s(t_0)\cap X_u(t_0)$, then $U(n+1,t_0)x\in \im
P^+_{n+1}$ and there is a $y\in \ker P^+_n$ such that $x=U(t_0, n)y$. 
Then $U(n+1,n)y=U(n+1,t_0)x\in \im P^+_{n+1}$. Thus,
$U(n+1,n)y=0$ and $y=0$ since $U(n+1,n):\ker P^+_n\to \ker P^+_{n+1}$ is an isomorphism. 
Thus, $X_s(t_0)\cap X_u(t_0)=\{0\}$. To prove that $X=X_s(t_0)\oplus
X_u(t_0)$, take an $x\in X$, and decompose $U(n+1,t_0)x=y_s+y_u$, $y_s\in \im P^+_{n+1}$, $y_u\in \ker P^+_{n+1}=X_u (n+1)$. Let $x_u$ denote the unique vector in
$X_u(t_0)$ such that $U(n+1,t_0)x_u=y_u$, and let $x_s=x-x_u$. Then $x_s\in X_s(t_0)$ since 
$U(n+1,t_0)x_s=y_s\in \im P^+_{n+1}$. Projections $P^+_t,t\geq 0$, with
$\im P^+_t=X_s(t)$, $\ker P^+_t=X_u(t)$ give the desired dichotomy.
The proof of the "if" part of the lemma  is straightforward. \hfill$\Box$

\section{Special Cases}\label{S5}
In this section we discuss several particular cases when the 
 statements of Theorems~\ref{main} and \ref{mainRef} 
allow certain simplifications, and 
indicate classes of  problems for which 
these theorems could be applied. We present the 
results only for $L_p=L_p(\mathbb{R};X)$, $p\in[1,\infty)$.
In this section all differential equations $u'(t)=A(t)u(t)$ with,
generally, unbounded operators $A(t)$,
 $t\in \mathbb{R}$,  are assumed to be well-posed in the following 
$W_p^1(\mathbb{R};X)$--sense
(cf. \cite[p.313]{Sch}):
(1) There exists a dense subset $\mathcal{D}\subset X$ such that
$\dom A(t)=\mathcal{D}$ for all $t\in\mathbb{R}$; and (2) There exists
a stongly continuous exponentially bounded evolution family
$\{U(t,\tau)\}_{t\ge\tau}$, $t,\tau\in\mathbb{R}$, on $X$ so that for all
$\tau\in\mathbb{R}$ and each $x_\tau\in\mathcal{D}$ 
the function $u(t)=U(t,\tau)x_\tau$, defined for $t\ge\tau$, takes values in
$\mathcal{D}$, belongs to 
the Sobolev space $W_p^1([\tau,\infty);X)$, and satisfies the differential
equation $u'(t)=A(t)u(t)$ for almost all $t\ge\tau\in\mathbb{R}$.

{\bf Mild and regular solutions.} 
The operator ${\bf G}$, described in Lemma~\ref{DefESG}, is the
generator of the evolution semigroup induced by the propagator 
$\{U(t,\tau)\}_{t\ge\tau}$ of 
the well posed differential equation $u'(t)=A(t)u(t)$, $t\in\mathbb{R}$. 
Therefore, ${\bf G}$ is a closed operator on $L_p(\mathbb{R};X)$,
$p\in [1,\infty)$. Also, $u:\mathbb{R}\to X$ is a mild solution of 
the inhomogeneous equation $u'(t)=A(t)u(t)+f(t)$, $t\in \mathbb{R}$, 
for $f\in L_p(\mathbb{R};X)$, provided $u\in \dom {\bf G}$ and ${\bf G}u=f$. 
Consider the operator $G=-d/dt+A(t)$ with the domain $\dom G$ given in 
\eqref{domG}. We say that
$u$ is a regular solution of the inhomogeneous equation 
provided $u\in \dom G$ and $Gu=f$. Note that for many classes of equations
(say, parabolic) mild solutions have additional regularity. If this is the
case, one might expect that ${\bf G}=G$. The latter equality is indeed
true provided, for instance, that the
inhomogeneous equation $u'(t)=A(t)u(t)+f(t)$ has $L_p$-maximal regularity,
a property established for a large variety of parabolic nonautonomous
problems, see \cite{LB,Sch} for further references.

Recall that, by \cite[Thm. 3.12]{ChLa99} and \cite[Prop.4.1]{Sch},
the set $\dom G$ from \eqref{domG} is a core for ${\bf G}$. 
Thus, if $G$ is closed then ${\bf G}=G$. As a result, we conclude
that if $G$ is a closed  operator 
on $L_p(\mathbb{R};X)$, $p\in [1,\infty)$, 
then Theorems \ref{main} and \ref{mainRef},
and all other results of this paper, are valid if the operator 
${\bf G}$ in their formulations is replaced by $G$.
We will not go into discussion
of the (quite delicate, see \cite[Sec.(c)]{Sch}) 
question when $G$ is closed, but merely
mention that ${\bf G}=G$ under the following simplest assumption:
\begin{equation}\label{bound}
  A:\mathbb R \mapsto 
 \mathcal{L}(X) \text { is piecewise continuous and } \sup_{t\in \mathbb{R}}\| A(t)\|<\infty.
\end{equation}
Indeed, in this case the propagator $\{ U(t,\tau
)\}_{t,\tau \in \mathbb{R}}$ is
differentiable in $\mathcal{L}(X)$. 
Then $u\in W^1_p(\mathbb{R};X)$ is a regular solution of
the inhomogeneous equation if and only if $u$ is a mild solution of this
equation. Therefore,  ${\bf G}=G$ 
for the operator $G=-d/dt+A(t)$ with $\dom
G=W^1_p(\mathbb{R};X)$, $p\in [1,\infty)$.

{\bf Compactness and node operators.}  In many cases studied in the literature the operator ${\bf G}$ (or $G$, defined in
\eqref{DifOp} with the domain \eqref{domG}) was proved to be Fredholm if and only if the 
corresponding evolution family 
(or the differential equation $u'(t)=A(t)u(t)$, $t\in \mathbb{R}$) 
has exponential dichotomies on $\mathbb{R}_+$ and $\mathbb{R}_-$, see, e.g.,
\cite[Thm.1.2]{BAG}, \cite[Thm.1.1]{HSS}, \cite[Lem.3.4]{Lin}, 
\cite[Lem.4.2]{Pa} and \cite{Pal}, \cite[Thm.2.6]{SS1}, 
 \cite[Thm.1.3]{Z}. Thus, in these papers condition (ii') in 
Theorem~\ref{mainRef} or, equivalently, see Lemma~\ref{REQ},
condition (ii) in Theorem~\ref{main} has been fulfilled automatically. 
A reason for this is explained in Lemma \ref{PJ16.11} below.
Indeed, under the assumptions imposed in the above
cited papers, or for the classes of the evolution families studied in these papers, the projectors $I-P^+_0$ and $I-P^-_0$ 
 happened to be of finite rank (and thus compact), or their difference
was compact. If, for instance, 
$U(t,\tau )$ are compact operators in $X$ for all $t > \tau$ in
$\mathbb{R}$, then the invertibility of 
their restrictions $U(t,\tau)|_{\ker P_{\tau}}$ acting from
 $\ker P_{\tau}$ to $\ker P_{t}$
(see (ii) in the definition of the exponential dichotomy) 
implies that $\ker P_{\tau}$  is finite dimensional. The more general {\em
$\alpha$-contractivity} condition on $U(t,\tau )$ also implies that $\ker P_{\tau}$ is  finite dimensional, see, e.g., \cite[p.
21]{SS94} and the literature cited therein. 
The following two examples, on the contrary, identify important
autonomous equations $u'(t)=Au(t)$ for which {\em both} stable and
unstable subspaces are infinite dimensional, see also \cite{SS2,SS1}.

\begin{exmp}\label{EJ1612.1} {\em ({\em Petrovskij-correct systems.})
 Let ${\bf p}(\xi )=[{\bf
p}_{kj}(\xi)]^K_{k,j=1}$, $\xi \in \mathbb{R}^d$, $d\geq 1$, be a $(K\times K)$ matrix whose entries are complex-valued polynomials ${\bf
p}_{ij}(\xi )=\sum_{|\alpha |\leq N_{kj}}a_\alpha \xi^\alpha$. Here we use the multiindex notation for $\alpha \in \mathbb{N}^d$, and
$a_\alpha
\in \mathbb{C}$ depend on $k$ and $j$. In $L_2(\mathbb{R}^d;\mathbb{C}^K)$ the operator $A={\bf p}(i\partial )$, $\partial =(\partial
_1,\ldots ,\partial _d)$, $i^2=-1$, is defined via Fourier transform, $A=\mathcal{F}^{-1}{\bf p} (\cdot )\mathcal{F}$, and is a general
(matrix) constant coefficient operator with the symbol ${\bf p}$.
We say that $A$ is {\em Petrovskij correct} if for some 
$\omega \in \mathbb{R}$  the spectrum $\sigma({\bf p}(\xi ))$ of the
matrix ${\bf p}(\xi )$ satisfies $\sigma ({\bf p}(\xi ))\subset \{z\in
\mathbb{C}:\re z\leq \omega \}$ for all $\xi \in \mathbb{R}^d$. If this is the case, then $A$ generates a strongly continuous semigroup on
$L_2(\mathbb{R}^d)$, where $\dom A$ is the Sobolev space of order $N=\max N_{kj}$. This semigroup is hyperbolic provided $\sigma ({\bf
p}(\xi ))$ is uniformly separated from $i\mathbb{R}$ for all $\xi \in \mathbb{R}^d$. Both stable and unstable spectral subspaces can be infinite
dimensional. A ``toy" $(2\times 2)$ matrix first order example is 
${\bf p}(\xi )=\diag [i\xi -a, i\xi +b]$, $\xi \in \mathbb{R}$, $a,b>0$,
where $\sigma (A)=(i\mathbb{R}-a) \cup (i\mathbb{R}+b)$.
For a study of dichotomy  of  hyperbolic systems 
with constant and close to constant coefficients see  
\cite{KS,SV} and the literature therein.} \hfill$\Diamond$\end{exmp}

\begin{exmp}\label{PerShr}{\em
({\em Schr\"{o}dinger operators with periodic potentials.}) Consider on $X=L_2(\mathbb{R};\mathbb{C})$ a
Schr\"{o}dinger operator $A=\frac{d^2}{dx^2}+V(x)$, $\dom
A=W_2^2(\mathbb{R};\mathbb{C})$, with a piecewise continuous real-valued
periodic potential $V$. By Theorem XIII.90 from \cite{ReedS} we know
that its spectrum
$\sigma(A)=\cup_{n=1}^\infty[\alpha_n,\beta_n]$ for
some $\beta_n\le\alpha_{n+1}$, and $\sigma(A)$ is absolutely continuous;
also, unless
$V$ is a constant, $\alpha_{n+1}\neq\beta_n$ for some $n$, that is, there are
gaps in $\sigma(A)$ (e.g., $\alpha_{n+1}\neq\beta_n$ for all $n\in\mathbb{N}$
for the Mathieu potential $V(x)=\mu\cos x$, $\mu\neq 0$). Thus, if
$0\in(\beta_n,\alpha_{n+1})$ for some $n$ then the equation
$u'(t)=Au(t)$ has an exponential dichotomy on $\mathbb{R}$
 with infinite dimensional stable and unstable subspaces.}
\hfill$\Diamond$\end{exmp}

\begin{lem}\label{PJ16.11} If $P^+_0$ and $P^-_0$ are projectors on 
a Banach space $X$, and $P^+_0-P^-_0$ is a compact operator, then the node
operator $N(0,0)=(I-P^+_0)|_{\ker P^-_0}:\ker P^-_0\to \ker P^+_0$ is 
Fredholm.\end{lem}

\begin{pf} A $(2\times2)$ matrix representation \eqref{1B.zero} 
of the Fredholm operator $L=I-(P^+_0-P^-_0)$ acting from 
$X=\im P^-_0\oplus \ker P^-_0$ to $X=\im P^+_0\oplus \ker P^+_0$ 
has the form
$L= \left[\begin{smallmatrix} P^+_0P^-_0 & 0 \\
 2(I-P^+_0)P^-_0 & N(0,0)\end{smallmatrix}\right]$,
where $N(0,0)=(I-P^+_0)(I-P^-_0):\ker P^-_0\to \ker P^+_0$. By (ii) in Lemma~\ref{LSp}, $\im N(0,0)$ is closed and $\dim \ker
N(0,0)<\infty$. Passing to the adjoints, $N(0,0)^*=[I -(P^-_0)^*(P^+_0-P^-_0)^*]|_{\ker (P^+_0)^*}$. Since $P^+_0-P^-_0$ is compact,
$\dim \ker N(0,0)^*<\infty$.\hfill$\Box$\end{pf}

The assumption of Lemma \ref{PJ16.11} is often used in
the literature on Morse theory in Hilbert spaces, in particular, 
for the study of Fredholm differential operator $G$
 on infinite-dimensional spaces in \cite{AM1} and  \cite{AM}.
To establish a link between  the current work and \cite{AM1,AM} 
assume, for a moment, that
$X$ is a Hilbert space, and $(P_W,P_V)$ is a pair of selfadjoint 
projections on subspaces $W$ and $V$ of $X$, respectively. 
The pair $(W, V)$ is called \emph {commensurable} 
 if the operator 
$P_W-P_V$ is compact, see \cite[Ch.2]{ABB}. It can be shown that if the pair $(W, V)$ is commensurable, then
the pair $(W, V^{\bot})$ is Fredholm, and
\begin{equation}\label{J18.2.1} 
\ind (W,V^{\bot})=\dim (W,V),
\end{equation}
where the {\em relative dimension}, $\dim (W,V)$, of subspaces $W$ and $V$ is defined by $\dim (W,V):=\dim (W\cap V^\bot)-\dim (W^\bot \cap
V)$, see~\cite[Sec. 2.2]{ABB}. Here, subspaces $W$ and $V$ are, in general, infinite dimensional. However, if $\dim W<\infty$ and $\dim
V<\infty$, then $\dim (W,V)=\dim W-\dim V$.
\begin{exmp}{\em
To illustrate the simple fact that not every Fredholm pair of subspaces is commensurable, let  
$P_{W}=\frac{1}{2} \left[\begin{smallmatrix}  I &  I\\
  I &  I \end{smallmatrix}\right]$ and  
$P_{V}=\left[\begin{smallmatrix} I & 0\\ 0 & 0 \end{smallmatrix}\right]$ be 
selfadjoint projections on the subspaces
${W}= \{\ x\oplus x: x \in H\}$
and ${V}=\{ x \oplus 0: x \in H \}$ of the orthogonal direct 
sum $X$ of two copies of an infinite dimensional Hilbert space $H$.
Then $P_W -P_V$ is not compact (since it is invertible), 
but $W + V^{\bot}=X$ and
$W \cap V^{\bot}= \{0\},$ and thus $(W,V^{\bot})$ is a Fredholm pair.}
\hfill$\Diamond$
\end{exmp}
If an evolution family $\{U(t,\tau )\}_{t\geq \tau}$ 
has exponential dichotomies $\{ P^+_t\}_{t\geq 0}$ and $\{
P^+_t\}_{t\leq 0}$ on $\mathbb{R}_+$, resp., on $\mathbb{R}_-$, 
then only the  subspaces $\im P^+_0$  (stable for $t\to\infty$)
and $\ker P^-_0$ (stable for $t\to-\infty$) are uniquely
determined, see e.g, \cite[Rem.IV.3.4]{DK} and \cite[Eqn.(3.20)]{SS2}. 
Thus, if $X$ is  a Hilbert space, we can assume 
in Propositions \ref{PJ183} and \ref{PJ18.4} below
that $P^+_0$ and $P^-_0$ are selfadjoint projections. 
Lemma~\ref{PJ16.11} and formula \eqref{J18.2.1} for $W=\ker P^{-}_0$ and $V=\ker P^+_0$
lead to the following abridged version of Theorem~\ref{mainRef}
that, nevertheless, covers many known results.
In particular, the index formula below gives the 
corresponding formulas from \cite{BAG}, \cite{Pal}, 
and is related to \cite[Theorem B]{AM1}
(see also Proposition \ref{PJ18.4} below).
\begin{prop}\label{PJ183}   Suppose that an evolution family 
$\{ U(t,\tau )\}_{t\geq \tau}$ on a Banach space 
$X$ has exponential dichotomies
$\{P^+_t\}_{t\geq 0}$ and $\{P^-_t\}_{t\leq 0}$ on $\mathbb{R}_+$ and $\mathbb{R}_-$ such 
that the operator $P^+_0-P^-_0$ is compact.
Then the following holds:
\begin{enumerate}\item[(a)]
 ${\bf G}$ is Fredholm on $L_p(\mathbb{R};X)$, 
$p\in [1,\infty )$; $\ind {\bf G}=\ind (\ker P^-_0 ,\im P^+_0)$; 
\item[(b)]
If, in addition, $X$ is
a Hilbert space and $P_0^{\pm}$ 
are selfadjoint projections, then $\ind {\bf G}=
\dim (\ker P^-_0, \ker P^+_0)$; 
\item[(c)] If, moreover,  $\dim \ker P^\pm_0<\infty$, 
then $\ind {\bf G}=\dim \ker P^-_0-\dim \ker P^+_0.$
\end{enumerate}
Conversely, if the operators $U(t,\tau )$, $t>\tau \in \mathbb{R}$, on a
reflexive Banach space $X$ are compact, and ${\bf G}$ is Fredholm, 
then there exist exponential dichotomies $\{ P^+_t\}_{t\geq 0}$ and $\{
P^-_t\}_{t\leq 0}$, and $\dim \ker P^\pm _0<\infty$. \end{prop}

\noindent{\bf Perturbations.}
 Consider a well-posed differential equation 
$u'(t)=A(t)u(t)$, $t\in \mathbb{R}$, with the propagator $\{ U_A(t,\tau
)\}_{t\geq \tau}$, $t,\tau \in\mathbb{R}$, and a perturbation 
$B:\mathbb{R}\to \mathcal{L}(X)$. 
We will impose the following assumptions\footnote{Apparently, the
assumption that $B(t),t\in \mathbb{R}$, are {\em bounded} operators could be relaxed to include wider classes of perturbations, cf.
\cite[Sec. 5.2.2]{ChLa99}, but will not pursue this here.}
on the perturbation:
\begin{enumerate}
\item[$(P_1)$] The function $t\mapsto B(t)x$ is  
continuous for each $x\in X$; 
\item[$(P_2)$] $\sup_{t\in \mathbb{R}}\| B(t)\| <\infty$; 
\item[$(P_3)$] the perturbed equation $u'(t)=[A(t)+B(t)]u(t)$ is 
well posed with the propagator $\{ U_{A+B}(t,\tau )\}_{t\geq \tau }$, 
$t,\tau \in\mathbb{R}$;
\item[$(P_4)$] $\lim_{|t|\to \infty}\| B(t)\|=0$;
\item[$(P_5)$] $B(t)$ is a compact operator for each $t\in \mathbb{R}$.
\end{enumerate}
We remark that assumption $(P_3)$ is not trivial in view of an example due to R. Phillips, see, e.g., \cite[Exmp. 2.3]{Sch}. 
Let ${\bf G}_A$ and ${\bf G}_{A+B}$ denote the generators of the evolution semigroups  induced
 by $\{ U_A(t,\tau )\}_{t\geq \tau}$ and $\{
U_{A+B}(t,\tau )\}_{t\geq \tau}$, respectively. Under assumptions
$(P_1)$--$(P_3)$ it can be shown that ${\bf G}_{A+B}={\bf G}_A+\mathcal{B}$, where
$\mathcal{B}\in\mathcal{L}(L_p(\mathbb{R};X))$ 
is defined by $(\mathcal{B}u)(t)=B(t)u(t)$, 
 a.e. $t\in \mathbb{R}$, cf.~\cite[Thm. 5.24]{ChLa99}.
Obviously, $\mathcal{B}$ may not be compact. As an example, consider $\mathcal{B}$ 
with $B(t)=\alpha(t) B,$ where $\alpha  \in C_0 (\mathbb R;\mathbb{R}),
 \alpha \ne 0,$ and $B$ is a compact operator such that
 $\sigma (B) \neq \{ 0\}$.
 Then $\sigma (\mathcal {B})=\{\alpha (t) :  t \in \mathbb R\} \cdot \sigma(B)$ 
is uncountable. 
\begin{prop}\label{PJ18.1} Suppose that $B$ satisfies
 assumptions $(P_1)-(P_5)$.
Then ${\bf G}_A$ and ${\bf G}_{A+B}$ are Fredholm on $L_p(\mathbb{R};X)$,
$p\in [1,\infty)$, simultaneously, and $\ind {\bf G}_A=\ind {\bf G}_{A+B}$.
\end{prop}

\begin{pf} Let $D_A$ and $D_{A+B}$ denote the difference operators on 
$\ell_p(\mathbb{Z};X)$, $p\in[1,\infty)$, induced by the evolution families
$\{U_A(t,\tau )\}_{t\geq \tau }$ and $\{ U_{A+B}(t,\tau )\}_{t\geq \tau}$ 
using \eqref{2.1}. By Theorem~\ref{FrGandD}, we need to show
that $D_A$ and $D_{A+B}$ are Fredholm at the same time with equal indexes. By the standard perturbation theory, the perturbed evolution
family $\{ U_{A+B}(t,\tau )\}_{t\geq \tau}$ satisfies a variation of constants
formula for all $t\geq \tau$. This formula, in particular, implies
$U_{A+B}(n+1,n)x=U_A(n+1,n)x+K_{n+1}x$, for all $x\in X$ and 
$n\in \mathbb{Z}$, where
$$K_{n+1}x=\int^{n+1}_{n}U_{A+B}(n+1,s)B(s)U_A(s,n)x\, ds.$$
Then $D_{A+B}-D_A={\mathcal {K}}$, where ${\mathcal {K}}:=\diag [ K_n ]_{n\in \mathbb{Z}}:
(x_n)_{n \in \mathbb{Z}}\mapsto (K_nx_n)_{n\in \mathbb{Z}}$. Since $B(s)\to 0$
as $|s|\to \infty$ in ${\mathcal {L}}(X)$, and the evolution families $\{ U_A(t,\tau )\}_{t\geq \tau}$ and $\{ U_{A+B}(t,\tau )\}_{t\geq \tau}$ are exponentially
bounded, we have $\lim_{|n|\to \infty}K_n=0$ in ${\mathcal {L}}(X)$.  Also, since operators $B(s)$,
$s\in \mathbb{R}$, are compact and the functions $f_n(\cdot)=U_{A+B}(n+1,\cdot)B(\cdot)U_A(\cdot,n)$ are strongly continuous on $[n, n+1],  n \in \mathbb Z,$ 
we conclude that $K_n$ is compact
in $X$ for each $n\in \mathbb{Z}$, see, e.g. \cite[p.525]{EnNa99}. Thus, ${\mathcal {K}}$ is 
compact in $\ell_p(\mathbb{Z};X)$  as a limit in 
${\mathcal {L}}(\ell_p(\mathbb{Z};X))$  of a sequence of compact operators. \hfill$\Box$ \end{pf}

\noindent{\bf Asymptotically constant coefficients.} 
Let $A$ be the  generator of a strongly continuous semigroup 
$\{ e^{tA}\}_{t\geq 0}$ on $X$.
The evolution family corresponding to the equation $u'(t)=Au(t)$ 
is given by $U(t,\tau )=e^{(t-\tau )A}$ for $t\geq \tau$ in $\mathbb{R}$.
Recall that a semigroup $\{ e^{tA}\}_{t\geq 0}$ is called {\em hyperbolic} on $X$ if 
there exists a projection $P_A$ such that
$e^{tA}P_A=P_Ae^{tA}$, $t\geq 0$, and that 
$\| e^{tA}|_{\im P_A}\| \leq Me^{-\alpha t}$, $t\geq 0$, $\alpha > 0,$ and the semigroup $\{ e^{tA}|_{\ker
P_A}\}_{t\geq 0}$ extends to a strongly continuous {\em group} 
$\{ e^{tA}|_{\ker P_A}\}_{t\in \mathbb{R}}$ on $\ker P_A$ such that $\|
e^{tA}|_{\ker P_A}\| \leq Me^{\alpha t}$, $t\leq 0$, see, e.g., \cite[p. 28]{ChLa99}. The semigroup $\{ e^{tA}\}_{t\geq 0}$ is hyperbolic
if and only if $\sigma (e^{tA})\cap \mathbb{T}=\emptyset$ for some (and hence
 for all) $t>0$. Then  $P_A$ is the spectral 
(Riesz) projection for $\{ e^{tA}\}_{t \ge 0}$ such
that  $\sigma (e^{tA}|_{\im P_A})=\sigma (e^{tA})\cap \{\lambda \in \mathbb{C}:|\lambda|<1\}$, 
see \cite[Lem.2.15]{ChLa99}.

\begin{lem}\label{CONSTH} Assume that for some $b\geq 0$ the evolution 
family $\{ e^{(t-\tau)A}\}_{t\geq \tau}$, $t,\tau \in \mathbb{R}$, has either an
exponential dichotomy $\{ P^+_t\}_{t\geq b}$ on $[b,+\infty)$,
 or an exponential dichotomy $\{ P^-_t\}_{t\leq -b}$ on $(-\infty ,-b]$. Then the semigroup $\{
e^{tA}\}_{t\geq 0}$ is hyperbolic on $X$.\end{lem}

\begin{pf} We will prove that $\sigma (e^A)\cap \mathbb{T}=\emptyset$ 
provided there is a dichotomy $\{ P^+_t\}_{t\geq b}$. First, we claim that 
$\| (I-e^A)x\|\geq
c\|x\|$ for some $c>0$ and all $x\in X$. By Lemma~\ref{LNN8.1}, for some $c>0$ we have $\| D^+_b{\bf x}\|_{\ell_p}\geq c\| {\bf x}\|_{\ell_p}$ for
all ${\bf x}\in \ell_p(\mathbb{Z}\cap [b+1,\infty);X)$. For each $x\in X$ and 
$\gamma >0$ define ${\bf x}=(e^{-\gamma n}x)_{n\geq b+1}$. 
Then $D^+_b{\bf x}=(e^{-\gamma (b+1)}x,
(e^{-\gamma (b+2)}-e^Ae^{-\gamma (b+1)})x,\ldots )$. 
A calculation shows that
\begin{equation*}\begin{split} \| D^+_b{\bf x}\|^p_{\ell_p}&
=e^{-\gamma p(b+1)}\{ \| x\|^p+\| (e^{-\gamma }-e^A)x\| ^p/(1-e^{-\gamma p})\}\\
&\geq c^p\| {\bf x}\|^p_{\ell_p}=c^pe^{-\gamma p(b+1)}\| x\|^p/(1-e^{-\gamma p}).\end{split}\end{equation*}
Thus, $(1-e^{-\gamma p})\|x\|^p+\|(e^{-\gamma}-e^A)x\|^p\geq c^p\|x\|^p$, 
and letting $\gamma \to 0$ the claim is proved. 
Rescaling $A\mapsto A-i\beta$, $\beta
\in \mathbb{R}$, shows that $\| (\lambda-e^A)x\|\geq c\|x\|$ 
for all $x\in X$ and $\lambda \in \mathbb{T}$. 
To finish the proof of the lemma, it suffices to show
that $\sigma_p((e^{A})^*)\cap \mathbb{T}=\emptyset$ for 
the point spectrum $\sigma_p(\cdot)$. Arguing by contradiction and 
using the Spectral Mapping Theorem for
the point spectrum (\cite[Sec.IV.3.b]{EnNa99}, and also see
\cite[Sec.IV.2.18]{EnNa99}), suppose that 
$A^*\xi =i\beta \xi$ for some $\beta \in \mathbb{R}$ and $\xi \in X^*$. 
Then $(e^{sA})^*\xi=e^{i\beta s}\xi$ for all $s\geq 0$. 
Using the dichotomy $\{ P^+_t\}_{t\geq b}$ and passing to the adjoints, 
for all $t\geq b$ we have
$(P^{+}_b)^*(e^{(t-b)A})^*=(e^{(t-b)A})^*(P^{+}_t)^*$, and the dichotomy 
estimates $\| (e^{(t-b)A})^*|_{\im (P^{+}_t)^*}\| \leq Me^{-\alpha (t-b)}$, 
$\|(e^{(t-b)A})^*|_{\ker (P^{+}_t)^*})^{-1}\| \leq Me^{-\alpha (t-b)}$. 
Denote $\xi_t=e^{-i\beta (t-b)}(e^{(t-b)A})^*\xi$, $t\geq b$. 
Identity $(e^{(t-b)A})^*\xi=e^{i\beta
(t-b)}\xi$ implies that $\xi =(P^{+}_b)^*\xi_t+(I-(P^{+}_b)^*)\xi_t$
for all $t\ge b$. 
By the stable dichotomy estimate $\| (P^{+}_b)^* \xi_t\| =
\|(e^{(t-b)A})^*(P^{+}_t)^*\xi \| \leq
Me^{-\alpha (t-b)}\sup_{t\geq b}\| (P^{+}_t)^*\|\| \xi \|$, 
and we have $\lim_{t\to \infty} (I-(P^{+}_b)^*)\xi_t=
\lim_{t\to\infty}[\xi -(P^{+}_b)^*\xi_t]=\xi \in \ker
(P^{+}_b)^*$ since $(I-(P^{+}_b)^*)\xi _t\in \ker (P^{+}_b)^*$. 
By the unstable dichotomy estimate,
$$\| \xi \|=\|(I-(P^{+}_b)^*)\xi _t\| 
=\|(e^{(t-b)A})^*(I-(P^{+}_t)^*)\xi \| 
\geq M^{-1}e^{\alpha (t-b)}\| (I-(P^{+}_t)^*)\xi \|,$$ 
and so $\lim_{t\to\infty}(I-(P^{+}_t)^*)\xi=0$. Using the
decomposition $\xi=(P^{+}_t)^*\xi+(I-(P^{+}_t)^*)\xi$, we have
$\xi =\lim_{t\to \infty}(P^{+}_t)^*\xi$.
Remark that $\im P^+_t=\im P^+_b$ for all $t\geq b$.
Indeed, using the dichotomy $\{P_t^+\}_{t\ge b}$, for each $t\ge b$ we infer:
\begin{equation*}\begin{split} 
\im P^+_t &=\{ x\in X:\| e^{A(s-t)}x\| 
\leq Me^{-\alpha (s-t)}\| x\|\text{ for all } s\geq t
\}\\
& =\{ x\in X:\| e^{A\tau}x\| \leq Me^{-\alpha \tau}\| x\| \text{ for all }
\tau\geq 0\}=
\im P^+_b.\end{split}\end{equation*} 
Since $\im (P^{+}_t)^*=(\im P_t^+)^*$, for all $t\ge b$
we thus have $\im (P^{+}_t)^*=\im (P^{+}_b)^*$.
Therefore, $(P^{+}_t)^*\xi \in \im (P^{+}_b)^*$ implies $\xi 
=\lim_{t\to \infty}(P^{+}_t)^*\xi\in \im
(P^{+}_b)^*$, and so $\xi =0$ since we have proved that 
$\xi\in\ker(P_b^+)^*\cap\im(P_b^+)^*$. 
Dichotomy $\{ P^-_t\}_{t\leq -b}$ is considered similarly.
\hfill$\Box$\end{pf}

\begin{cor}\label{PCC} Let $A$ be the generator of a strongly 
continuous semigroup on a reflexive Banach space $X$.
Then the following assertions are equivalent:
\begin{enumerate}\item[(1)] ${\bf G}_A$ is Fredholm on 
$L_p(\mathbb{R};X)$, $p\in [1,\infty )$;
\item[(2)] ${\bf G}_A$ is invertible on $L_p(\mathbb{R};X)$, $p\in [1,\infty )$;
\item[(3)] $\sigma (e^{tA})\cap \mathbb{T}=\emptyset$ for all $t>0$.\end{enumerate}
\end{cor}

\begin{pf} The equivalence (2)$\Leftrightarrow$(3) is contained 
in \cite[Thm. 3.13]{ChLa99}. To prove (1)$\Rightarrow$(3), apply
Theorem~\ref{main}. By this theorem, (1) implies the existence 
of an exponential dichotomy 
$\{ P_t^+\}_{t\geq b}$ on $[b,\infty)$ for
the evolution family $\{ e^{(t-\tau )A}\}_{t\geq \tau}$. 
By Lemma \ref{CONSTH} the semigroup $\{ e^{tA}\}_{t\geq 0}$ is hyperbolic.
\hfill$\Box$\end{pf}
Next, consider a perturbed differential equation $u'(t)=[A+B(t)]u(t)$, 
$t\in \mathbb{R}$. If assumption $(P_4)$ holds then this equation
is asymptotically autonomous (for a recent work on asymptotically autonomous
parabolic equations see also \cite{BC,DGL,Sch2,Sch3}). 
\begin{lem}\label{LJ15.1}  Suppose that assumptions  $(P_1)$--$(P_4)$ hold.
Assume that for some
$b\geq 0$ the evolution family $\{U_{A+B}(t,\tau )\}_{t\geq \tau}$ 
for $u'(t)=[A+B(t)]u(t)$, $t\in \mathbb{R}$, has 
either an exponential
dichotomy $\{ P^+_t\}_{t\geq b}$ on $[b,\infty )$, or  an exponential
dichotomy $\{ P^-_t\}_{t \leq -b}$ on $(-\infty,-b].$
Then the semigroup $\{ e^{tA}\}_{t\geq 0}$ is hyperbolic.\end{lem}

\begin{pf}  Suppose that the evolution family 
$\{U_{A+B}(t,\tau )\}_{t\geq \tau}$ has 
 an exponential
dichotomy $\{ P^+_t\}_{t\geq b}$ on $[b,\infty )$
 with the dichotomy constants $\alpha$, $M$. 
Since $B(t)\to 0$ in ${\mathcal {L}}(X)$ as $t\to\infty$ 
by  assumption ($P_4$), for
each $\epsilon \in (0,\alpha (2M)^{-1})$ there exists a $T=T(\epsilon )\geq b$ such that $\sup \{ \| B(t)\| :t\geq T\}<\epsilon$. For $t\in
\mathbb{R}$ we set $\tilde{P}_t=P^+_t$ if 
$t\geq T$ and $\tilde{P}_t=P^+_T$ if $t<T$. 
Also, we define a strongly continuous exponentially
bounded evolution family 
$\{ \tilde{U}_{A+B}(t,\tau )\}_{t\geq \tau}$, $t,\tau\in \mathbb{R}$, 
a continuation of $\{ U_{A+B}(t,\tau
)\}_{t\geq \tau \geq T}$, by 
\begin{equation}\label{dA+B}
 \tilde{U}_{A+B}(t,\tau )=\begin{cases} U_{A+B}(t,\tau )& \text{  for  } t\geq \tau \geq T,\\
 U_{A+B}(t,T)e^{\alpha (T-\tau )(I-2P^+_T)}& \text {  for  } t\geq T\geq \tau,\\
e^{\alpha (t-\tau)(I-2P^+_T)}& \text{  for  }
T\geq t\geq \tau, \end{cases}
\end{equation} 
cf. \cite[p.109]{BAG}.  Since $e^{\alpha (t-\tau)(I-2P^+_T)}=
e^{-\alpha (t-\tau)}P^+_T+e^{\alpha (t-\tau)}(I-P^+_T)$, it is
easy to check that $\{ \tilde{P}_t\}_{t\in \mathbb{R}}$ is an exponential dichotomy for $\{ \tilde{U}_{A+B}(t,\tau )\}_{t\geq \tau}$ on
$\mathbb{R}$ with the same dichotomy constants $\alpha ,M$. By \cite[Thm. 3.13]{ChLa99}, 
the generator ${\bf \tilde{G}}_{A+B}$ of the evolution
semigroup on $L_p(\mathbb{R};X)$ induced by $\{ \tilde{U}_{A+B}(t,\tau )\}_{t\geq \tau}$ is invertible, and, moreover, $\|
({\bf \tilde{G}}_{A+B})^{-1}\|_{\mathcal{L}(L_p(\mathbb{R};X))}
\leq 2M\alpha^{-1}$, see, e.g. \cite[p. 105]{ChLa99}. 
Extend the evolution family
$\{ e^{(t-\tau )A}\}_{t\geq \tau \geq T}$ as follows: 
\begin{equation}\label{dA}
 \tilde{U}_A(t,\tau )=\begin{cases} e^{(t-\tau )A}& \text{  for  } t\geq \tau \geq T,\\
e^{A(t-T)}e^{\alpha (T-\tau )(I-2P^+_T)}& \text{  for  } t\geq T\geq \tau,\\ 
e^{\alpha (t-\tau)(I-2P^+_T)}& \text{  for  } T\geq t\geq \tau.
\end{cases}
\end{equation}
Define $\tilde{B}:\mathbb{R}\to \mathcal{L}(X)$ by setting $\tilde{B}(t)=B(t)$ for $t\geq T$ and
$\tilde{B}(t)=0$ for $t<T$, and define $\tilde{\mathcal{B}}\in \mathcal{L}(L_p(\mathbb{R};X))$ by
$\tilde{\mathcal{B}}u(t)=\tilde{B}(t)u(t)$, $t\in \mathbb{R}$. Then ${\bf \tilde{G}}_{A+B}={\bf \tilde{G}}_A+\tilde{\mathcal{B}}$, where
${\bf \tilde{G}}_A$ is the generator of the evolution semigroup on $L_p(\mathbb{R};X)$ induced by the evolution family
$\{\tilde{U}_A(t,\tau )\}_{t\geq \tau}$. By the choice of $T$, 
$$\| \tilde{\mathcal{B}}\|_{\mathcal{L}(L_p(\mathbb{R};X))}=\sup_{t\geq T}\| B(t)\|<\epsilon <\alpha (2M)^{-1}\leq (\| ({\bf
\tilde{G}}_{A+B})^{-1}\|_{\mathcal{L}(L_p(\mathbb{R};X))})^{-1}.$$
Thus, ${\bf \tilde{G}}_A$ is invertible on $L_p(\mathbb{R}_+;X)$ since ${\bf \tilde{G}}_{A+B}$ 
is invertible. By \cite[Thm. 3.13]{ChLa99}, the
evolution family
$\{\tilde{U}_A(t,\tau )\}_{t\geq \tau }$ has an exponential 
dichotomy on $\mathbb{R}$, hence, on $[T,\infty)$, and thus $\{ e^{tA}\}_{t\geq
0}$ is hyperbolic by Lemma \ref{CONSTH} with $b=T$.
The case of exponential dichotomy on $(-\infty, b]$ 
is considered similarly.\hfill$\Box$ \end{pf}
\begin{prop}\label{PJ157.2} Assume that $A$ is the generator of a 
strongly continuous semigroup on a reflexive Banach space $X$,
and assumptions $(P_1)$--$(P_5)$ hold for a perturbation
$B:{\mathbb{R}}\to{\mathcal{L}}(X)$.
Then ${\bf G}_{A+B}$
is Fredholm on $L_p(\mathbb{R};X)$, $p\in [1,\infty )$, if and only if the semigroup $\{ e^{tA}\}_{t\geq 0}$ is hyperbolic. Moreover, $\ind
{\bf G}_{A+B}=0$.\end{prop}
\begin{pf} If ${\bf G}_{A+B}$ is Fredholm, 
then $\{ U_{A+B}(t,\tau )\}_{t\geq \tau}$ has an exponential dichotomy 
on $\mathbb{R}_+$ by
Theorem~\ref{mainRef}. By Lemma~\ref{LJ15.1},
 $\{ e^{tA}\}_{t\geq 0}$ is hyperbolic. Conversely, if $\{ e^{tA}\}_{t\geq 0}$ is
hyperbolic then ${\bf G}_A$ is invertible on 
$L_p$, see Corollary~\ref{PCC}. By Proposition \ref{PJ18.1} 
${\bf G}_{A+B}$ is Fredholm and $\ind{\bf G}_{A+B}=0$.
\hfill$\Box$\end{pf}
There is an alternative proof of Lemma \ref{LJ15.1}, appropriate
for $C_0(\mathbb{R};X)$, that uses difference
operators, cf. the proof of Proposition \ref{PJ18.1}. This proof is based
on the fact that if $D_{A+B}$ and $D_A$ are the difference operators  \eqref{2.1}
 induced by the evolution families defined in \eqref{dA+B} and \eqref{dA},
respectively, then $\|D_{A+B}-D_A\|$ is small provided
$\|{\tilde{\mathcal{B}}}\|$ is small.
Also, because of Lemma~\ref{LJ15.1}, assumption $(P_5)$ on $B$ was used only in
the proof of the ``only if" part of Proposition~\ref{PJ157.2}. Thus, for any $B\in C_0(\mathbb{R};\mathcal{L}(X))$, if ${\bf G}_{A+B}$ is
Fredholm, then $\{ e^{tA}\}_{t\geq 0}$ is hyperbolic.

{\bf Asymptotically piecewise constant coefficients.} 
Let $A_+$ and $A_-$ be the  generators of strongly continuous semigroups $\{
e^{tA_+}\}_{t\geq 0}$ and $\{ e^{tA_-}\}_{t\geq 0}$ on $X$, respectively. Assume that $\dom A_+=\dom A_-$, and let
\begin{equation}\label{J15.34} A_0(t)=A_+\text{ for } t\geq 0\text{ and }A_0(t)=A_-\text{ for } t<0.\end{equation}
Then the differential equation $u'(t)=A_0(t)u(t)$, $t\in \mathbb{R}$, 
is well-posed in the $W_p^1$--sense with a 
propagator $\{ U(t,\tau )\}_{t\geq \tau}$, $t,\tau \in \mathbb{R}$, defined as follows:
\begin{equation}
U(t,\tau)=\begin{cases}e^{(t-\tau )A_+}& \text{ for } t\geq \tau \geq 0,\\
e^{tA_+}e^{-\tau A_-}& \text{ for } t\geq 0\geq \tau,\\
e^{(t-\tau )A_-}&\text{ for } 0\geq t\geq \tau .\end{cases}\label{J153.1}
\end{equation}
The invertibility of ${\bf G}_{A_0}$ with bounded operators $A_\pm$ has been
studied in \cite{Ch}.
\begin{prop}\label{PPCC}  Let $A_0$ be defined by \eqref{J15.34},
where $\dom A_+=\dom A_-$.
The operator ${\bf G}_{A_0}$ is Fredholm on $L_p(\mathbb{R};X)$,
$p\in [1,\infty)$, if and only if 
\begin{enumerate}\item[(1)] The semigroups $\{ e^{tA_+}\}_{t\geq 0}$ 
and $\{ e^{tA_-}\}_{t\geq 0}$ are hyperbolic on $X$ with
the spectral projections $P_{A_+}$ and $P_{A_-}$, respectively;
\item[(2)] The node operator $N(0,0)=(I-P_{A_+})|_{\ker P_{A_-}}:
\ker P_{A_-}\to \ker P_{A_+}$ is Fredholm.\end{enumerate}
Moreover, $\dim\ker{\bf G}_{A_0}=\dim\ker N(0,0)$,
$\codim\im{\bf G}_{A_0}=\codim\im N(0,0)$, and
$\ind {\bf G}_{A_0}=\ind N(0,0)$.\end{prop}

\begin{pf} If (1) and (2) hold then ${\bf G}_{A_0}$ 
is Fredholm and the index formula is valid by the ``if" part of 
Theorem~\ref{mainRef}. If
${\bf G}_{A_0}$ is Fredholm, then by the ``only if" part 
of Theorem~\ref{mainRef}, 
there exist dichotomies $\{P^+_t\}_{t\geq 0}$ and $\{
P^-_t\}_{t\leq 0}$ for the evolution family 
$\{ U(t,\tau )\}_{t\geq \tau}$ defined in \eqref{J153.1}.
By Lemma \ref{CONSTH}, the semigroups $\{e^{tA_{\pm}}\}_{t\ge0}$ are
hyperbolic, and we may set $P^+_t=P_{A_+}$ and $P^-_t=P_{A_-}$. 
This proves (1). Assertion (2) holds by the implication
\eqref{GFr} $\Rightarrow$ (ii') in Theorem \ref{mainRef}, and Lemma \ref{REQ}.
\hfill$\Box$\end{pf}

Next, consider $A(t)=A_0(t)+B(t)$ with $B$
satisfying assumptions  $(P_1)$--$(P_3)$, 
and let ${\bf G}_{A_0+B}$ and ${\bf G}_{A_0}$ denote the generators of the evolution semigroups induced by
the propagators of the differential equations 
$u'(t)=[A_0(t)+B(t)]u(t)$ and $u'(t)=A_0(t)u(t)$, respectively.
Recall that if $\sigma(A)\cap i\mathbb{R}=\emptyset$ then $P_A$ denotes the
spectral projection such that $\sigma(A|_{\im P_A})=\sigma(A)\cap\{\lambda \in\mathbb{C}: \re \lambda <0\}$.

\begin{prop}\label{PJ15.8.1} Assume that $A_+$ and $A_-$,
$\dom A_+=\dom A_-$,  are the generators of strongly continuous semigroups on 
a reflexive Banach space $X$, and
$B$ satisfies  assumptions {{$(P_1)$--$(P_5)$}}.
The operator ${\bf G}_{A_0+B}$ is Fredholm if and only if the semigroups $\{
e^{tA_+}\}_{t\geq 0}$ and $\{ e^{tA_-}\}_{t\geq 0}$ are hyperbolic with 
the spectral projections $P_{A_+}$ and $P_{A_-}$, and the pair of subspaces
$(\ker P_{A_-},\im P_{A_+})$ is Fredholm. Moreover, $\ind {\bf G}_{A_0+B}=\ind 
(\ker P_{A_-},\im P_{A_+})$.\end{prop}

\begin{pf} By Proposition \ref{PJ18.1}, ${\bf G}_{A_0+B}$ and ${\bf G}_{A_0}$ are Fredholm
at the same time, and their indexes are equal. 
The rest follows from Proposition~\ref{PPCC} and Lemma \ref{REQ}.\hfill$\Box$\end{pf}

\begin{cor}\label{CJ22.1} 
Let $X$ be a separable Hilbert space.
Assume that $A_+$ and $A_-$ are selfadjoint operators with 
compact resolvent, and $\dom A_+=\dom A_-$. Let $A_0$ be defined
as in \eqref{J15.34}. Suppose that $B:\mathbb{R}\to \mathcal{L}(X)$ satisfies assumptions $(P_1)-(P_5)$, and that $B(t)$ is a selfadjoint operator for each $t\in
\mathbb{R}$. Then ${\bf G}_{A_0+B}$ is Fredholm if and only if $A_+$ and $A_-$ are invertible. Moreover, $\ind {\bf G}_{A_0+B}$ is equal to the spectral flow for
the family $A(t)=A_0(t)+B(t)$, $t\in \mathbb{R}$.\end{cor}

Recall, that the spectral flow for the family $\{A(t)\}_{t\in \mathbb{R}}$ of selfadjoint operators with 
compact resolvent represents the net change in the number
of negative eigenvalues of $A(t)$ as $t$ changes from $-\infty $ 
to $+\infty$, see e.g. \cite{RS} or \cite[Sec.8.16]{Mel}. 
In the situation described in 
Corollary~\ref{CJ22.1} we thus {\em define} the
spectral flow as $\dim \ker P_{A_-}-\dim \ker P_{A_+}$, cf. \cite{DGL}.
Note that $A(t)$ has compact resolvent for all $t\in\mathbb{R}$.

\begin{pf} By the spectral mapping theorem 
$\sigma (e^{tA})\backslash \{ 0\}=\exp t\sigma (A)$, $t>0$, for 
selfadjoint operators \cite[Thm.IV.3.10]{EnNa99}, the operator $A_\pm $ is
invertible if and only if the semigroup $\{ e^{tA_\pm }\}_{t\geq 0}$ 
is hyperbolic. Since $A_+$ and $A_-$ have compact resolvents, 
$\ker P_{A-}$ and $\ker P_{A_+}$ are finite dimensional,
and $P_{A_+}-P_{A_-}$
is compact. Thus, subspaces $\ker P_{A_-}$ and $\ker P_{A_+}$ are commensurable, and, by Lemma~\ref{PJ16.11}, the node operator $N(0,0)$ is Fredholm. So, by
Lemma~\ref{REQ} the pair of subspaces $(\ker P_{A_-},\im P_{A_+})$ is Fredholm. 
Using formula \eqref{J18.2.1}
for $W=\ker P_{A_-}$ and $V=(\im P_{A_+})^\bot$, we conclude that $\ind (\ker P_{A_-}, \im P_{A_+})=\dim \ker P_{A_-}-\dim \ker P_{A_+}$. An application of
Proposition~\ref{PJ15.8.1} concludes the proof.\hfill$\Box$\end{pf}

\noindent{\bf Bounded coefficients.} 
Assume that \eqref{bound} holds, and recall that ${\bf G}_A=G_A$.
 Let $\{ U(t,\tau )\}_{t,\tau
\in \mathbb{R}}$ denote the propagator for $u'(t)=A(t)u(t)$, 
$t\in \mathbb{R}$. 
If $\{ U(t,\tau )\}_{t,\tau \in \mathbb{R}}$ has exponential 
dichotomies $\{ P^+_t\}_{t\geq 0}$ and $\{ P^-_t\}_{t\leq 0}$ on
$\mathbb{R}_+$ and $\mathbb{R}_-$, then the stable, $W^s_A,$ and unstable, $W^u_A,$ subspaces for $A$ 
 can be described as follows:
\begin{align*}W^s_A&=\{x\in X:\lim_{t\to \infty}U(t,0)x=0\}=\im P^+_0,\\ W^u_A&=\{ x\in X:\lim_{t\to -\infty}U(t,0)x=0\}=\ker
P^-_0.\end{align*} 
\begin{prop}\label{PJ17.1} Assume that  $A$ satisfies \eqref{bound}.  
Then the operator $G_A$  
is Fredholm if and only if the following holds:\,\,\, 
(a) There exist exponential dichotomies $\{ P^+_t\}_{t\geq 0}$ 
and $\{P^-_t\}_{t\leq 0}$ on $\mathbb{R}_+$ and $\mathbb{R}_-$ 
for $\{ U(t,\tau )\}_{t,\tau\in\mathbb{R}}$; 
and (b) The pair of subspaces
$(W^s_A,W^u_A)$ is Fredholm. Moreover, $\ind G=\ind (W^s_A,W^u_A).$ 
\end{prop}

This follows from Theorem \ref{mainRef}. Further, if the limits $A_+=\lim_{t\to \infty}A(t)$ and $A_-=\lim_{t\to -\infty}A(t)$ exist in $\mathcal{L}(X)$, and $\sigma (A_\pm )\cap
i\mathbb{R}=\emptyset$, then the operator family $\{ A(t)\}_{t\in \mathbb{R}}$ is called an  {\em asymptotically hyperbolic} path; see,
e.g., \cite{AM1}. Under the additional assumption that $\{ A(t)\}_{t\in \mathbb{R}}$ is asymptotically hyperbolic,
Proposition~\ref{PJ17.1} has been proved in \cite[Thm. D]{AM1}. 
Our results show, however, that if the limits $A_+$ and $A_-$ exist and the operator $G_A$ is
Fredholm, then $\sigma (A_\pm )\cap i\mathbb{R}=\emptyset$. Indeed, since $G_A$ is Fredholm, 
Theorem~\ref{main} implies the existence of
dichotomies $\{ P^+_t\}_{t\geq b}$ and $\{ P^-_t\}_{t\leq a}$ for some $a\leq b$. 
Using the assumption that $A_\pm =\lim_{t\to \pm
\infty}A(t)$ exist in $\mathcal{L}(X)$, this, in turn, implies that $\sigma (e^{tA_\pm})\cap \mathbb{T}=\emptyset$, $t>0$, see
Lemma~\ref{LJ15.1}.
Further, for $A_\pm \in \mathcal{L}(X)$ define $A_0$ as in \eqref{J15.34}, and consider a compact-valued perturbation $B:\mathbb{R}\to
\mathcal{L}(X)$ that satisfies assumptions $(P_1)-(P_5)$. Proposition~\ref{PJ15.8.1} and formula \eqref{J18.2.1} give the following
improvement of \cite[Thm.B]{AM1}, 
where the ``if'' part of Proposition \ref{PJ18.4}  has
been proved. 

\begin{prop}\label{PJ18.4}If $A(t)=A_0(t)+B(t), t \in \mathbb R,$ where $A_0$ is  given by \eqref{J15.34} with 
$A_\pm \in \mathcal{L}(X)$, and $B$ takes compact values and vanishes at $\pm \infty$, then $G_A$ is Fredholm 
on $L_p(\mathbb{R};X)$, $p\in
[1,\infty )$, if and only if $\sigma (A_\pm )\cap i\mathbb{R}=\emptyset$ 
and the pair of the spectral subspaces $(\im P_{A_+},\ker
P_{A_-})$ for $A_+$ and $A_-$ is Fredholm. Moreover, $\ind G_A=\ind (\ker P_{A_-},\im P_{A_+})$. 
If $X$ is a Hilbert space and, in
addition, $A_+-A_-$ is a compact operator,  
and $P_{A_{\pm}}$ are selfadjoint projections, 
 then $\ind G_A=\dim (\ker P_{A_-},\ker P_{A_+})=\dim (\im
P_{A_+},\im P_{A_-}).$ \end{prop}

{\bf Connections to Morse Theory.} A need to study Fredholm properties and 
the index of the operator $G$ naturally arises in infinite dimensional Morse theory,
see~\cite{ABB,AM} and the literature therein. If $X=\mathbb{R}^d$ and $v$ is a (heteroclinic) solution 
of the equation $v'(t)=f(v(t))$ connecting two hyperbolic
stagnation points, $x_-=\lim_{t\to -\infty} v(t)$ and 
$x_+=\lim_{t\to \infty}v(t)$, then the linearization along 
$v$ gives rise to the operator $Gu=-u'+A(t)u$, where
$A(t)=Df(v(t))$, $t\in \mathbb{R}$, and $Df$ is the differential. 
If $f$ is a gradient vector field, that
is, $f=-DF$ for a Morse functional $F:X\to \mathbb{R}$ 
(such that $D^2 F(x)$ is hyperbolic at all critical points $x$ of $F$),
then $A(\pm\infty)=-D^2F(x_{\pm})$, and the number $\dim \ker P_{-D^2F(x_\pm)}=
\dim \ker P_{A(\pm\infty)}$ is called the Morse index of the critical
point $x_\pm$. It 
is well-known that $\ind G=\dim \ker P_{A(-\infty)}-\dim \ker
P_{A(+\infty)}$, see, e.g., \cite[Thm.2.1]{RS}. If $X$ is an infinite dimensional Hilbert space then Morse functionals of particular interest are of the form
$F(x)=\frac{1}{2}\langle Ax,x\rangle +b(x)$ since they appear
in the study of Hamiltonian systems, wave equations, and some elliptic
systems, see \cite{ABB,AM}. Here $A$ is a selfadjoint operator and the Hessian $D^2F(x)=A+D^2b(x)$, where $D^2b(x)$ is a compact operator
on $X$ for each $x\in X$. If, as above, $v$ is a heteroclinic trajectory connecting (hyperbolic) critical points, then the linearization along $v$ gives the
operator $Gu=-u'+A(t)u$, where $A(t)=A+B(t)$, $B(t)=D^2b(v(t))$, 
$t\in \overline{\mathbb{R}}$. In the infinite dimensional
situation just outlined, the Morse theory has been developed in
\cite{AM}. Note, that the results of the current section (see Proposition~\ref{PJ15.8.1} 
and Corollary~\ref{CJ22.1}) show that the hyperbolicity of the operators
 $D^2F(x_{\pm})$ is, in fact,
necessary for the operator $G$ to be Fredholm. Moreover, it appears 
that Theorem~\ref{mainRef} is applicable for more
general Morse functionals. 
In this case, the exponential dichotomies on 
$\mathbb{R}_\pm $ in this 
theorem seem to be a correct
generalization of the asymptotic hyperbolicity.

\noindent{\bf Travelling waves.} Applications of the finite dimensional 
Dichotomy Theorem in the theory of travelling waves are important and
well-understood, see \cite{Sa} and the literature therein. 
We briefly sketch a 
simple generalization of the setup in \cite{Sa},
suitable for applications of the infinite
dimensional version of this theorem given in the current paper 
(cf. \cite{SS2} and \cite[pp.89--91]{SS1}). Let $Y$ be a Banach space, $\mathcal{N}:Y\to Y$ be a
differentiable nonlinear map, ${\bf p}(\cdot )$ be a polynomial with constant coefficients, $u:\mathbb{R}_+\times \mathbb{R}\to Y$. Consider a
nonlinear equation
\begin{equation}\label{NLu} \partial_tu={\bf p}(\partial _x)u+\mathcal{N}(u),\quad t\in
\mathbb{R}_+,\quad x\in \mathbb{R}.\end{equation}
A typical situation occurs when 
$u=u(t,x,y)$, $y\in\mathbb{R}^d$, and
$Y=L_2(\mathbb{R}^d)$, so that $u(t,\cdot,\cdot)
\in L_2(\mathbb{R}\times
\mathbb{R}^d)=L_2(\mathbb{R};L_2(\mathbb{R}^d))$ and $u(t,x,\cdot )\in L_2(\mathbb{R}^d)$. 
In our general setting, passing to the moving frame $\xi =x-ct$,
$c\neq 0$, $v(t,\xi )=u(t,\xi +ct)$, $\xi \in \mathbb{R}$, we have that 
$u$ satisfies \eqref{NLu} if and only if $v$ satisfies
\begin{equation}\partial _tv={\bf p}(\partial_\xi )v+c\partial_\xi v+\mathcal{N}(v),\quad t\in \mathbb{R}_+,\quad \xi \in
\mathbb{R}.\label{NLv}\end{equation} 
A function $\mathbf{q}=\mathbf{q}_c(\xi )$, $\mathbf{q}:\mathbb{R}\to Y$, is called a {\em
travelling wave} for \eqref{NLu} if $\mathbf{q}$ is a $t$-independent solution of \eqref{NLv}, that is, if 
${\bf p}(\partial_\xi)\mathbf{q}+c\partial_\xi
\mathbf{q}+\mathcal{N}(\mathbf{q})=0$. Assume that the latter (nonlinear) equation 
has a solution. A linearization of \eqref{NLv} about $\mathbf{q}$ gives rise to an operator
\begin{equation}\label{OpL.1} Lw:={\bf p}(\partial _\xi )w+
c\partial_\xi w+D\mathcal{N}(\mathbf{q}(\xi ))w,\quad w=w(\xi )\in Y,
\quad\xi\in\mathbb{R}.\end{equation}
In a ``general" semilinear case we might have $\mathcal{N}(u)=Nu+F(u)$, 
where $N$ is any generator of a strongly continuous semigroup on $Y$. 
If, in addition, $DF(0)=0$, 
$\mathbf{q}(\xi )\to 0$ as $|\xi |\to \infty$, 
and for each $\xi \in \mathbb R$ the operator 
$B(\xi )=DF(\mathbf{q}(\xi ))$ 
is a compact operator on $Y$, then
our perturbation results are applicable. 
Finally, we note that the eigenvalue problem $Lw=\lambda w$ for $L$ in
\eqref{OpL.1} is a higher order nonautonomous ordinary differential equation in $Y$ and, as such, could be rewritten as a first order
equation $u'(\xi )=A(\xi )u(\xi )$, where $A(\xi ),\xi \in \mathbb{R}$, depends on $\lambda$ and, generally, is an unbounded differential
operator on a suitable Banach space $X=Y\oplus \ldots \oplus Y$. Thus, 
the spectrum of $L$ is related to the Fredholm properties of the
operator ${\bf G}_A$ 
induced by $A$ which are described in the current paper.

\end{document}